\documentclass[12pt]{article}

\usepackage{amssymb}
\usepackage{amsmath}
\usepackage{amscd}
\usepackage{latexsym}
\usepackage{graphicx}

\usepackage{cite}

\newcommand{\be}{\begin{equation}}
\newcommand{\ee}{\end{equation}}
\newcommand{\Dlt}{\Delta}
\newcommand{\dlt}{\delta}
\newcommand{\prt}{\partial}
\newcommand{\br}{{\bf r}}

\newcommand{\bt}{\beta}
\newcommand{\vp}{\varphi}
\newcommand{\ep}{\varepsilon}
\newcommand{\al}{\alpha}
\newcommand{\ra}{\rightarrow}
\newcommand{\sgm}{\sigma}

\newcommand{\gm}{\gamma}
\newcommand{\om}{\omega}

\newcommand{\Gm}{\Gamma}

\newcommand{\lbd}{\lambda}

\begin{document}

\begin{center}

{\Large{\bf From Asymptotic Series to Self-Similar Approximants} \\ [5mm]

V.I. Yukalov$^{1,2}$ and E.P. Yukalova$^{3}$ }  \\ [3mm]

{\it
$^1$Bogolubov Laboratory of Theoretical Physics, \\
Joint Institute for Nuclear Research, Dubna 141980, Russia \\ [2mm]

$^2$Instituto de Fisica de S\~ao Carlos, Universidade de S\~ao Paulo, \\
CP 369, S\~ao Carlos 13560-970, S\~ao Paulo, Brazil \\ [2mm]

$^3$Laboratory of Information Technologies, \\
Joint Institute for Nuclear Research, Dubna 141980, Russia } \\ [3mm]

{\bf E-mails}: {\it yukalov@theor.jinr.ru}, ~~ {\it yukalova@theor.jinr.ru}

\end{center}

\vskip 1cm

\begin{abstract}
The review presents the development of an approach of constructing 
approximate solutions to complicated physics problems, starting from asymptotic 
series, through optimized perturbation theory, to self-similar approximation 
theory. The close interrelation of underlying ideas of these theories is emphasized. 
Applications of the developed approach are illustrated by typical examples 
demonstrating that it combines simplicity with good accuracy. 
\end{abstract}

\vskip 2cm
{\parindent=0pt
{\bf Keyword}: asymptotic perturbation theory; optimized perturbation theory;
self-similar approximation theory; optimized approximants; root approximants; 
nested approximants; exponential approximants; factor approximants, combined 
approximants; data extrapolation; diff-log approximants; critical phenomena }

\newpage

\section*{Contents}

{\parindent=0pt
{\bf 1}. Introduction

\vskip 1mm
{\bf 2}. Asymptotic Expansions

\vskip 1mm
{\bf 3}. Sequence Transformations
\vskip 1mm
   \hspace{3mm} {\bf 3.1} Pad\'{e} Approximants
\vskip 1mm
   \hspace{3mm} {\bf 3.2} Borel Summation

\vskip 1mm
{\bf 4}. Optimized Perturbation Theory
\vskip 1mm
   \hspace{3mm} {\bf 4.1} Initial Approximation
\vskip 1mm
   \hspace{3mm} {\bf 4.2} Change of Variables
\vskip 1mm
   \hspace{3mm} {\bf 4.3} Sequence Transformations

\vskip 1mm
{\bf 5}. Statistical Physics

\vskip 1mm
{\bf 6}. Optimization Conditions

\vskip 1mm
{\bf 7}. Thermodynamic Potential

\vskip 1mm
{\bf 8}. Eigenvalue Problem

\vskip 1mm
{\bf 9}. Nonlinear Schr\"{o}dinger Equation

\vskip 1mm
{\bf 10}. Hamiltonian Envelopes
\vskip 1mm
   \hspace{4mm} {\bf 10.1} General Idea
\vskip 1mm   
   \hspace{4mm} {\bf 10.2} Power-Law Potentials
\vskip 1mm
   \hspace{4mm} {\bf 10.3} Inverse Power-Law Potentials
\vskip 1mm
   \hspace{4mm} {\bf 10.4} Logarithmic Potential

\vskip 1mm
{\bf 11}. Optimized Expansions: Summary 
\vskip 1mm
   \hspace{4mm} {\bf 11.1} Expansion over Dummy Parameters 
\vskip 1mm   
   \hspace{4mm} {\bf 11.2} Scaling Relations: Partition Function
\vskip 1mm
   \hspace{4mm} {\bf 11.3} Scaling Relations: Anharmonic Oscillator
\vskip 1mm
   \hspace{4mm} {\bf 11.4} Optimized Expansion: Partition Function
\vskip 1mm
   \hspace{4mm} {\bf 11.5} Optimized Expansion: Anharmonic Oscillator

\vskip 1mm
{\bf 12}. Order-Dependent Mapping
\vskip 1mm
   \hspace{4mm} {\bf 12.1} Change of Variables 
\vskip 1mm   
   \hspace{4mm} {\bf 12.2} Partition Function
\vskip 1mm
   \hspace{4mm} {\bf 12.3} Anharmonic Oscillator

\vskip 1mm
{\bf 13}. Variational Expansions

\vskip 1mm
{\bf 14}. Control Functions and Control Parameters  

\vskip 1mm
{\bf 15}. Self-Similar Approximation Theory 

\vskip 1mm
{\bf 16}. Embedding Cascade into Flow

\vskip 1mm
{\bf 17}. Stability Conditions

\vskip 1mm
{\bf 18}. Free Energy

\vskip 1mm
{\bf 19}. Fractal Transform

\vskip 1mm
{\bf 20}. Self-Similar Root Approximants

\vskip 1mm
{\bf 21}. Self-Similar Nested Approximants

\vskip 1mm
{\bf 22}. Self-Similar Exponential Approximants

\vskip 1mm
{\bf 23}. Self-Similar Factor Approximants

\vskip 1mm
{\bf 24}. Self-Similar Combined Approximants
\vskip 1mm
   \hspace{4mm} {\bf 24.1} Different Types of Approximants
\vskip 1mm
   \hspace{4mm} {\bf 24.2} Self-Similar Pad\'{e} Approximants
\vskip 1mm
   \hspace{4mm} {\bf 24.3} Self-Similar Borel Summation

\vskip 1mm
{\bf 25}. Self-Similar Data Extrapolation

\vskip 1mm
{\bf 26}. Self-Similar Diff-Log Approximants

\vskip 1mm
{\bf 27}. Critical Phenomena
\vskip 1mm
   \hspace{4mm} {\bf 27.1} Critical Point at Infinity
\vskip 1mm
   \hspace{4mm} {\bf 27.2} Finite Critical Point 
 
\vskip 1mm 
{\bf 28}. Non-Power-Law Behavior
\vskip 1mm
   \hspace{4mm} {\bf 28.1} Exponential Behavior
\vskip 1mm
   \hspace{4mm} {\bf 28.2} Logarithmic Behavior

\vskip 1mm
{\bf 29}. Critical Temperature Shift

\vskip 1mm
{\bf 30}. Critical Exponents

\vskip 1mm
{\bf 31}. Conclusion  }

\section{Introduction}

The standard way of treating realistic physical problems, described by complicated 
equations, relies on approximate solutions of the latter, since the occurrence of
exact solutions is rather an exception. The most often used method is a kind of 
perturbation theory based on expansions in powers of some small parameters. This 
way encounters two typical obstacles: the absence of small parameters and divergence 
of resulting perturbative series. To overcome these difficulties, different methods
of constructing approximate solutions have been suggested. 

In this review, we demonstrate how, starting from asymptotic series, there appear
general ideas of improving the series convergence and how these ideas lead to the 
development of powerful methods of optimized perturbation theory and self-similar
approximation theory.

\section{Asymptotic Expansions}            

Let us be interested in finding a real function $f(x)$ of a real variable $x$. A 
generalization to complex-valued functions and variables can be straightforwardly 
done by considering several real functions and variables. The case of a real function 
and variable is less cumbersome and allows for the easier explanation of the main 
ideas. Suppose that the function $f(x)$ is a solution of very complicated equations 
that cannot be solved exactly and allow only for finding an approximate solution for 
the asymptotically small variable $x \ra 0$ in the form
\be
\label{1}
f(x) \simeq f_k(x) \qquad (x \ra 0 ) \;   .
\ee
There can happen the following cases.

\vskip 2mm
(i) {\it Expansion over a small variable}:
\be
\label{2}
f_k(x) = f_0(x) \left( 1 + \sum_{n=1}^k a_n x^n \right) \; ,
\ee
where the prefactor $f_0(x)$ is a given function. The expansion is asymptotic in 
the sense of Poincar\'{e} \cite{Poincare_1,Dingle_2}, since
$$
\left| \; \frac{a_{n+1} x^{n+1}}{a_n x^n } \; \right| ~ \ra ~ 0 \qquad 
( x \ra 0) \;  ,
$$
with $a_n$ assumed to be nonzero. 

(ii) {\it Expansion over a small function}: 
\be
\label{3}
f_k(x) = f_0(x) \left[ 1 + \sum_{n=1}^k a_n \vp^n(x) \right] \; ,
\ee
when the function $\varphi(x)$ tends to zero as $x\ra 0$ so that
$$
\left| \; \frac{a_{n+1} \vp^{n+1}(x)}{a_n \vp^n(x) } \; \right| ~ \ra ~ 0 
\qquad  ( x \ra 0) \;   .
$$

(iii) {\it Expansion over an asymptotic sequence}:
\be
\label{4}
 f_k(x) = f_0(x) \left[ 1 + \sum_{n=1}^k a_n \vp_n(x) \right] \;  ,
\ee
such that
$$
\left| \; \frac{a_{n+1} \vp_{n+1}(x)}{a_n \vp_n(x) } \; \right| ~ \ra ~ 0 
\qquad  ( x \ra 0) \;     .
$$
        
(iv) {\it Generalized asymptotic expansion}:
\be
\label{5}
f_k(x) = f_0(x) \left[ 1 + \sum_{n=1}^k a_n(x) \vp_n(x) \right]   ,
\ee
where the coefficients $a_n(x)$ depend on the variable $x$ and $\{\varphi_n(x)\}$ 
is an asymptotic sequence, such that
$$
\left| \; \frac{a_{n+1}(x)\vp_{n+1}(x)}{a_n(x)\vp_n(x) } \; \right| ~ \ra ~ 0 
\qquad  ( x \ra 0) \;    .
$$
This type of expansion occurs in the Lindstedt-Poincar\'{e} technique 
\cite{Poincare_1,Nayfeh_3,Malley_4} and in the Krylov-Bogolubov averaging method
\cite{Krylov_5,Bogolubov_6,Grebennikov_7,Sanders_8}.
 
(v) {\it Expansion over a dummy parameter}:
\be
\label{6}
 f_k(x) = f_0(x) \left[ 1 + \sum_{n=1}^k a_n(x) \ep^n \right] \;   .
\ee
Here the value of interest corresponds to the limit $\varepsilon = 1$, while the
series is treated as asymptotic with respect to $\varepsilon \ra 0$, hence
$$
\left| \; \frac{a_{n+1}(x)\ep^{n+1}}{a_n(x)\ep^n } \; \right| ~ \ra ~ 0 
\qquad  ( \ep \ra 0) \;    .
$$
The introduction of dummy parameters is often used in perturbation theory, for 
instance in the Euler summation method, N\"{o}rlund method, and in the Abel 
method \cite{Hardy_9}. 

Dummy parameters appear when one considers a physical system characterized by 
a Hamiltonian (or Lagrangian) $H$, while starting the consideration with an 
approximate Hamiltonian $H_0$, so that one has  
$$
 H_\ep = H_0 + ( H - H_0 ) \ep  \qquad ( \ep \ra 1) \; .
$$
Then perturbation theory with respect to $H-H_0$ yields a series in powers of 
$\ep$. Different iteration procedures also can be treated as expansions in powers 
of dummy parameters. 

Sometimes perturbation theory with respect to a dummy parameter is termed 
nonperturbative, keeping in mind that it is not a perturbation theory with respect 
to some other physical parameter, say a coupling parameter. Of course this misuse 
of terminology is confusing, mathematically incorrect, and linguistically awkward. 
Therefore it is mathematically correct to call perturbation theory with respect to 
any parameter perturbation theory.

\section{Sequence Transformations}

Asymptotic series are usually divergent. To assign to a divergent series an 
effective limit, one involves different resummation methods employing sequence 
transformations \cite{Dyke_10}. The most often used are the Pad\'e approximation 
and Borel summation.

\subsection{Pad\'{e} Approximants}

The method of Pad\'{e} approximants sums the series
\be
\label{7}
f_k(x) = \sum_{n=0}^k a_n x^n
\ee
by means of rational fractions 
\be
\label{8}
P_{M/N}(x) = \frac{a_0 + \sum_{n=1}^M b_n x^n}{1 + \sum_{n=1}^N c_n x^n} 
\qquad ( M + N = k ) \;  ,
\ee
with the coefficients $b_n$ and $c_n$ expressed through $a_n$ from the requirement 
of coincidence of the asymptotic expansions
\be
\label{9}
 P_{M/N}(x) \simeq f_k(x) \qquad ( x \ra 0 ) \;  .
\ee
  
As is evident from their structure, the Pad\'{e} approximants provide the 
best approximation for rational functions. However, in general they have several 
deficiencies. First of all, they are not uniquely defined, in the sense that for 
a series of order $k$ there are $C_k^2 + 2$ different Pad\'{e} approximants 
$P_{M/N}$, with $M + N = k$, where
$$
 C_k^n = \frac{k!}{(k-n)!n!} = \frac{k(k-1)(k-2)\ldots(k-n+1)}{n!} \;  ,
$$
and there is no uniquely defined general prescription of which of them to choose. 
Often, one takes the diagonal approximants $P_{N/N}$, with $2N = k$. However, 
these are not necessarily the most accurate \cite{Baker_11}. Second, there is 
the annoying problem of the appearance of spurious poles. 

Third, when the sought function, at small $x$ behaves as in expansion (\ref{7}),
but at large $x$ it may have the power-law behavior $x^\bt$ that should be predicted
from the extrapolation of the small-variable expansion, then this extrapolation to 
a large variable $x\gg 1$ cannot in principle be done if $\bt$ is not known or 
irrational. Let us stress that here we keep in mind the extrapolation problem from
the knowledge of only the small-variable expansion and the absence of knowledge
on the behavior of the sought function at large $x$. This case should not be confused
with the interpolation problem employing the method of two-point Pad\'e approximants,
when both expansions at small as well as at large variables are available \cite{Honda_119}. 
  
Finally, the convergence of the Pad\'e approximants is not a simple problem 
\cite{Baker_11,Baker_12}, especially when one looks for a summation of a series 
representing a function that is not known. In the latter case, one tries to observe what 
is called apparent numerical convergence which may be absent.    

As an example of a problem that is not Pad\'e summable \cite{Bender_118,Simon_13}, 
it is possible to mention the series arising in perturbation theory for the 
eigenvalues of the Hamiltonian
$$
H = -\;\frac{1}{2} \;\frac{d^2}{dx^2} + \frac{1}{2} \; x^2 + g x^m \;  ,
$$
where $x \in (-\infty, \infty)$, $g > 0$, and $m \geq 8$.

\subsection{Borel Summation}

The series (\ref{7}) can be Borel summed by representing it as the Laplace integral
\be
\label{10}
 f_k^B(x)  = \int_0^\infty e^{-t} B_k(tx) \; dt  
\ee
of the Borel transform
\be
\label{11}
 B_k(t) \equiv \sum_{n=0}^k \; \frac{a_n}{n!}\; t^n \;  .
\ee
This procedure is regular, since if series (\ref{7}) converges, then
$$
f_k(x) = \sum_{n=0}^k a_n x^n = \sum_{n=0}^k \; \frac{a_n}{n!}\; x^n 
\int_0^\infty e^{-t} t^n \; dt = 
$$
$$
=
\int_0^\infty e^{-t} \sum_{n=0}^k \; \frac{a_n}{n!}\; (tx)^n \; dt = 
\int_0^\infty e^{-t} B_k(tx)\; dt = f_k^B(x) \; .
$$
Conditions of Borel summability are given by the Watson theorem \cite{Hardy_9}, 
according to which a series (\ref{7}) is Borel summable if it represents a function 
analytic in a region and in that region the coefficients satisfy the inequality
$|a_n| \leq C^n n!$ for all orders $n$.

The problem in this method arises because the sought function is usually unknown, 
hence its analytic properties also are not known, and the behavior of  the 
coefficients $a_n$ for large orders $n$ is rarely available. When the initial
series is convergent, its Borel transform is also convergent and the integration and 
the summation in the above formula can be interchanged. However, when the initial 
series is divergent, the interchange of the integration and summation is not allowed.
One has, first, to realize a resummation of the Borel transform and after this to 
perform the integration. 

There are series that cannot be Borel summed. As an example, we can mention a model 
of a disordered quenched system \cite{Bray_14} with the Hamiltonian
$$
 H(g,\vp,\xi) = ( 1 + \xi) \vp^2 + g \vp^4 \;  ,
$$
in which $\vp\in(-\infty,\infty)$, $\xi\in(-\infty,\infty)$, and $g>0$, so that 
the free energy, as a function of the coupling parameter, is
$$
 f(g) = - \int_{-\infty}^\infty \ln \; Z(g,\xi) 
\exp\left( -\; \frac{\xi^2}{\sgm}\right) \; \frac{d\xi}{\sqrt{\pi\sgm} } \; ,
$$
where the statistical sum reads as
$$
Z(g,\xi) = \int_{-\infty}^\infty \exp\{\; -H(g,\vp,\xi)\; \} \;
\frac{d\vp}{\sqrt{\pi} } \; .
$$
By analytic means and by direct computation of $200$ terms in the perturbation 
expansion for the free energy, it is shown \cite{Bray_14} that the series is not 
Borel summable, since the resulting terms do not converge to any limit.  
  
Sometimes the apparent numerical convergence can be achieved by using the Pad\'e
approximation for the Borel transform under the Laplace integral, which is termed 
the Pad\'e-Borel summation.

\section{Optimized Perturbation Theory}

The mentioned methods of constructing approximate solutions tell us that there 
are three main ways that could improve the convergence of the resulting series. 
These are: (i) the choice of an appropriate initial approximation; (ii) change of 
variables, and (iii) series transformation. However the pivotal question arises: 
How to optimize these choices.

The idea of optimizing system performance comes from optimal control theory for 
dynamical systems \cite{Lewis_15}. Similarly to dynamical systems, the optimization 
in perturbation theory implies the introduction of control functions in order to 
achieve series convergence, as was advanced in Refs. 
\cite{Yukalov_16,Yukalov_17,Yukalov_18} and employed for describing anharmonic 
crystals \cite{Yukalov_17,Yukalov_18,Yukalov_19,Yukalov_20,Yukalov_21,Yukalov_22} 
and the theory of melting \cite{Yukalov_23}. Perturbation theory, complimented by 
control functions governing the series convergence is called {\it optimized 
perturbation theory}.   

The introduction of control functions means the reorganization of a divergent 
series into a convergent one. Formally, this can be represented as the operation
\be
\label{12}
  \hat R[\; u_k \;]\{\; f_k(x)\; \} = \{\; F_k(x,u_k) \; \} 
\ee
converting an initial series into a new one containing control functions $u_k(x)$. 
Then the optimized approximants are
\be
\label{13}
 \overline f_k(x) = F_k(x,u_k(x)) \;  .
\ee
The optimization conditions define the control functions in such a way that to 
make the new series $\{F_k(x,u_k(x))\}$ convergent, because of which this method 
is named {\it optimized perturbation theory}. The general approach to formulating 
optimization conditions is expounded in the review articles 
\cite{Yukalov_24,Yukalov_25}, and some particular methods are discussed in Refs. 
\cite{Dineykhan_26,Sissakian_27,Feranchuk_28}. Control functions can be implanted 
into perturbation theory in different ways. The main methods are described below.

\subsection{Initial Approximation}

Each perturbation theory or iterative procedure starts with an initial 
approximation. It is possible to accept as an initial approximation not a fixed 
form but an expression allowing for variations. For concreteness, assume we are 
considering a problem characterized by a Hamiltonian $H$ containing a coupling 
parameter $g$. Looking for the eigenvalues of the Hamiltonian, using perturbation 
theory with respect to the coupling, we come to a divergent series
\be
\label{14}
 E_k(g) = \sum_{n=1}^k c_n g^n \qquad ( g \ra 0 ) \;  .
\ee
As an initial approximating Hamiltonian, we can take a form $H_0(u)$ containing 
trial parameters. For brevity, we write here one parameter $u$. Then we define 
the Hamiltonian 
\be
\label{15}
H_\ep = H_0(u) + \ep [ \; H - H_0(u) \; ] \qquad (\ep \ra 1) \;   .
\ee
To find the eigenvalues of the Hamiltonian, we can resort to perturbation theory 
in powers of the dummy parameter $\varepsilon$, yielding
\be
\label{16}
E_k(g,u) = \sum_{n=1}^k c_n(g,u) \ep^n \;   .
\ee
Setting $\varepsilon=1$ and defining control functions $u_k(x)$ from optimization 
conditions results in the optimized approximants 
\be
\label{17}
 \overline E_k(g) = E_k(g,u_k(g) ) \;  .
\ee
The explicit way of defining control functions and particular examples will be 
described in the following sections.

\subsection{Change of Variables}

Control functions can be implanted through the change of variables. Suppose we 
consider a series
\be
\label{18}
f_k(x) = \sum_{n=0}^k a_n x^n  \; .
\ee
Accomplishing the change of the variable
\be
\label{19}
x = x(z,u) \; , \qquad z = z(x,u) \;   ,
\ee
we come to the functions $f_k(x(z,u))$. Expanding the latter in powers of the new 
variable $z$, up to the order $k$, gives
\be
\label{20}
f_k(x(z,u)) = \sum_{n=0}^k b_n(u) z^n  \qquad ( z \ra 0 ) \;   .
\ee
In terms of the initial variable, this implies
\be
\label{21}
F_k(x,u) = \sum_{n=0}^k b_n(u) z^n(x,u)  \;   .
\ee
Defining control functions $u_k(x)$ yields the optimized approximants (\ref{13}).

When the variable $x$ varies between zero and infinity, sometimes it is convenient
to resort to the change of variables mapping the interval $[0,\infty)$ to the 
interval $(-\infty, 1]$, passing to a variable $y$,
\be
\label{22}
x = \frac{u}{(1-y)^\om} = x(y,u,\om) \; ,
\ee
where $u>0$ and $\omega>0$ are control parameters \cite{Kleinert_29,Yukalov_30}. 
The inverse change of variables is
\be
\label{23}
 y = 1 - \left( \frac{u}{x} \right)^{1/\om} = y(x,u,\om) \;  .
\ee
The series (\ref{18}) becomes
\be
\label{24}
f_k(x(y,u,\om)) = \sum_{n=0}^k 
a_n \; \left[\; \frac{u}{(1-y)^\om}\; \right]^n  \;  .
\ee
Expanding this in powers of $y$, we obtain
\be
\label{25}
 F_k(x,u,\om) = \sum_{n=0}^k b_n(u,\om) \; [\; y(x,u,\om) \; ]^n \;  .
\ee
Defining control functions $u_k = u_k(x)$ and $\omega_k = \omega_k(x)$ gives 
the optimized approximants
\be
\label{26}
\overline f_k(x) = F_k(x,u_k(x),\om_k(x) ) \; .
\ee
Other changes of variables can be found in review \cite{Yukalov_25}.

\subsection{Sequence Transformations}

Control functions can also be implanted by transforming the terms of the given 
series by means of some transformation,
\be
\label{27}
 \hat T[\;u\;] f_k(x) = F_k(x,u_k) \; .
\ee
Defining control functions $u_k(x)$ gives $\overline{F}_k(x,u_k(x))$. Accomplishing 
the inverse transformation results in the optimized approximant
\be
\label{28}
 \overline f_k(x) = \hat T^{-1}[\;u\;]  \overline F_k(x,u_k(x) ) \;  .
\ee

As an example, let us consider the fractal transform \cite{Yukalov_24} that we 
shall need in what follows,
\be
\label{29}
\hat T[\;s\;] f_k(x) =  x^s f_k(x) = F_k(x,s) \; .
\ee 
For this transform, the scaling relation is valid:
\be
\label{30}
 \frac{F_k(\lbd x,s)}{F_k(x,s) } = 
\frac{f_k(\lbd x)}{f_k(x)} \; \lbd^s \;  .
\ee
The scaling power $s$ plays the role of a control parameter through which control 
functions $s_k(x)$ can be introduced \cite{Yukalov_31,Gluzman_32,Yukalov_33}.

\section{Statistical Physics}

In the problems of statistical physics, before calculating observable quantities, 
one has to find probabilistic characteristics of the system. This can be either 
probability distributions, or correlation functions, or Green functions. So, 
first, one needs to develop a procedure for finding approximations for these 
characteristics, and then to calculate the related approximations for observable 
quantities. Here we exemplify this procedure for the case of a system described 
by means of Green functions 
\cite{Yukalov_16,Yukalov_17,Yukalov_18,Yukalov_19,Yukalov_20,Yukalov_21}. 

Let us consider Green functions for a quantum statistical system with particle 
interactions measured by a coupling parameter $g$. The single-particle Green 
function (propagator) satisfies the Dyson equation that can be schematically 
represented as
\be
\label{31}
 G(g) = G_0 + G_0 \;\Sigma(G(g))\; G(g) \; ,
\ee
where $G_0$ is an approximate propagator and $\Sigma(G)$ is self-energy 
\cite{Kadanoff_34,Yukalov_35}. 

Usually, one takes for the initial approximation $G_0$ the propagator of 
noninteracting (free) particles, whose self-energy is zero. Then, iterating 
the Dyson equation, one gets the relation
\be
\label{32}
 G_{k+1}(g) = G_0 + G_0 \;\Sigma_k(G_k(g))\; G_k(g) \;   ,
\ee
which is a series in powers of the coupling parameter $g$. Respectively the 
sequence of the approximate propagators $\{G_k(g)\}$ can be used for calculating 
observable quantities
\be
\label{33}
A_k(g) = \sum_{n=0}^k c_n g^n 
\ee
that are given by a series in powers of $g$. This is an asymptotic series with 
respect to the coupling parameter $g \ra 0$, which as a rule is divergent for 
any finite $g$. 

Instead, it is possible to take for the initial approximation an approximate 
propagator $G_0(u)$ containing a control parameter $u$. This parameter can, for 
instance, enter through an external potential \cite{Yukalov_36} corresponding to 
the self-energy $\Sigma_0$. Then the Dyson equation reads as
\be
\label{34}
G(g) = G_0(u) + G_0(u) \; [\; \Sigma(G(g)) - \Sigma_0(G_0(u)) \; ]\; G(g) \; .
\ee
Iterating this equation \cite{Yukalov_37} yields the approximations for the 
propagator
\be
\label{35}
G_{k+1}(g,u) = G_0(u) + 
G_0(u) \; [\; \Sigma_k(G_k(g,u)) - \Sigma_0(G_0(u)) \; ]\; G_k(g,u) \;  .
\ee
This iterative procedure is equivalent to the expansion in powers of a dummy 
parameter.
  
Being dependent on the control parameter $u$, the propagators $G_k(g,u)$ generate 
the observable quantities $A_k(g,u)$ also depending on this parameter. Defining 
control functions $u_k(g)$ results in the optimized approximants
\be
\label{36}
\overline A_k(g) = A_k(g,u_k(g))
\ee
for observable quantities.

\section{Optimization Conditions}

The above sections explain how to incorporate control parameters into the 
sequence of approximants that, after defining control functions, become 
optimized approximants. Now it is necessary to provide a recipe for defining 
control functions.

By their meaning, control functions have to govern the convergence of the 
sequence of approximants. The Cauchy criterion tells us that a sequence 
$\{F_k(x,u_k)\}$ converges if and only if, for any $\ep>0$, there exists a 
number $k_\varepsilon$ such that
\be
\label{37}
| \; F_{k+p}(x,u_{k+p})- F_k(x,u_k) \; | < \ep
\ee
for all $k> k_\ep$ and $p>0$. 

In optimal control theory \cite{Lewis_15}, control functions are defined as the 
minimizers of a cost functional. Considering the convergence of a sequence, it 
is natural to introduce the convergence cost functional \cite{Yukalov_24}
\be
\label{38}
 C[\; u \;] = \frac{1}{2} \sum_k C^2(F_{k+p},F_k) \; ,
\ee
in which the Cauchy difference is defined,  
\be
\label{39}
 C(F_{k+p},F_k)  \equiv F_{k+p}(x,u_{k+p})- F_k(x,u_k)   \; .
\ee
To minimize the convergence cost functional implies the minimization of the 
Cauchy difference with respect to control functions, 
\be
\label{40}
\min_u |\; C(F_{k+p},F_k) \; | = 
\min_u |\; F_{k+p}(x,u_{k+p})- F_k(x,u_k) \; | \;   ,
\ee
for all $k \geq 0$ and $p \geq 0$.  

In order to derive from this condition explicit equations for control functions, 
we need to accomplish some rearrangements. If the Cauchy difference is small, this 
means that it is possible to assume that $u_{k+p}$ is close to $u_k$ and $F_{k+p}$ 
is close to $F_k$. Then we can expand the first term of the Cauchy difference in 
the Taylor series with respect to $u_{k+p}$ in the vicinity of $u_k$, which gives
\be
\label{41}
F_{k+p}(x,u_{k+p}) = \sum_{n=0}^\infty \frac{1}{n!} \; 
\frac{\prt^n F_{k+p}(x,u_k)}{\prt u_k^n} \; ( u_{k+p} - u_k )^n \;   .
\ee
Let us treat $F_{k+p}$ as a function of the discrete variable $p$, which allows 
us to expand this function in the discrete Taylor series
\be
\label{42}
 F_{k+p}(x,u_k) = \sum_{m=0}^\infty \frac{1}{m!} \; \Delta_p^m F_k(x,u_k)\; , 
\ee
where a finite difference of $m$-th order is
\be
\label{43}
\Delta_p^m F_k = 
\sum_{j=0}^m (-1)^{m-j} \; \frac{m!}{j!\; (m-j)!} \; F_{k+jp} \;  .
\ee
As examples of finite differences, we can mention
$$
\Delta_p^0 F_k =  F_k \; , 
\qquad 
\Delta_p^1 F_k \equiv \Dlt_p F_k = F_{k+p} - F_k \; ,
\qquad 
\Delta_p^2 F_k =  F_{k+2p} - 2F_{k+p} + F_k  \;  .
$$
Thus the first term in the Cauchy difference can be represented as
\be
\label{44}
F_{k+p}(x,u_{k+p} ) = 
\sum_{m,n=0}^\infty \frac{(u_{k+p}-u_k)^n}{m! \; n!} \;
\frac{\prt^n}{\prt u_k^n}\; \Dlt_p^m F_k(x,u_k)\;   .
\ee

Keeping in the right-hand side of representation (\ref{44}) a finite number of 
terms results in the explicit optimization conditions. The zero order is not 
sufficient for obtaining optimization conditions, since in this order
$$
F_{k+p}(x,u_{k+p} ) \cong F_k(x,u_k) \; ,
$$
hence the Cauchy difference is automatically zero, 
$$
C(F_{k+p},F_k) \cong 0 \; .
$$

In the first order, we have
\be   
\label{45}
F_{k+p}(x,u_{k+p}) \cong F_{k+p}(x,u_k) + 
(u_{k+p} - u_k) \; \frac{\prt}{\prt u_k} \; F_k(x,u_k) \; ,
\ee
which gives the Cauchy difference
\be
\label{46}
 C(F_{k+p},F_k) = F_{k+p}(x,u_k) - F_k(x,u_k) + 
(u_{k+p} - u_k) \; \frac{\prt}{\prt u_k} \; F_k(x,u_k) \; .
\ee
The minimization of the latter with respect to control functions implies
$$
 \min_u |\; C(F_{k+p},F_k) \; | \leq 
\min_u |\; F_{k+p}(x,u_k) - F_k(x,u_k) \; | +
$$
\be
\label{47}
+
\min_u \left| \; (u_{k+p} - u_k ) \; \frac{\prt}{\prt u_k} \; F_k(x,u_k) \; 
\right| \; .
\ee
 
Minimizing the first part in the right-hand side of expression (\ref{47}), we 
get the {\it minimal-difference condition}
\be
\label{48}
 \min_u |\; F_{k+p}(x,u_k) - F_k(x,u_k) \; |  
\ee
for the control functions $u_k = u_k(x)$. The ultimate form of this condition 
is the equality
\be
\label{49}
 F_{k+p}(x,u_k) - F_k(x,u_k) = 0 \;  .
\ee

The minimization of the second part of the right-hand side of expression (\ref{47}) 
leads to the {\it minimal-derivative condition}
\be
\label{50}
\min_u \left|\; (u_{k+p} - u_k ) \; \frac{\prt}{\prt u_k} \; F_k(x,u_k) \; 
\right| \;   .
\ee
The minimum of condition (\ref{50}) is made zero by setting
\be
\label{51}
\frac{\prt}{\prt u_k} \; F_k(x,u_k) = 0 \;   .
\ee
When this equation has no solution for the control function $u_k$, it is 
straightforward to either set
\be
\label{52}
u_{k+p} = u_k \qquad 
\left( \frac{\prt}{\prt u_k} \; F_k(x,u_k) \neq 0 \right) \;  ,
\ee
or to look for the minimum of the derivative
\be
\label{53}
\min_u \left|\; \frac{\prt}{\prt u_k} \; F_k(x,u_k) \; \right| \qquad 
( u_{k+p} \neq u_k) \; .
\ee

In this way, control functions are defined by one of the above optimization 
conditions. It is admissible to consider higher orders of expression (\ref{44}) 
obtaining higher orders of optimization conditions \cite{Yukalov_25}. 

Control functions can also be defined if some additional information on the 
sought function $f(x)$ is available. For instance, when the asymptotic behavior 
of $f(x)$, as $x\ra x_0$, is known, where
\be
\label{54}
 f(x) \simeq f_{as}(x) \qquad ( x \ra x_0 ) \; ,
\ee
then the control functions $u_k(x)$ can be defined from the {\it asymptotic 
condition}
\be
\label{55}
F_k(x,u_k) = \hat T[\; u\; ]\; f_k(x) \simeq \hat T[\; u\; ]\; f_{as}(x)
\qquad ( x \ra x_0 ) \;  .
\ee

\section{Thermodynamic Potential}

As an illustration of using the optimized perturbation theory, let us consider the 
thermodynamic potential 
\be
\label{56}
f(g) = - \ln\; Z(g)
\ee
of the so-called zero-dimensional anharmonic oscillator model with the statistical sum
\be
\label{57}
Z(g) = 
\frac{1}{\sqrt{\pi}} \int_{-\infty}^\infty \exp(- H[\;\vp\;]) \; d\vp
\ee
and the Hamiltonian
\be
\label{58}
 H[\;\vp\;] = \vp^2 + g \vp^4 \qquad ( g > 0 ) \; .
\ee

Taking for the initial approximation the quadratic Hamiltonian
\be
\label{59}
H_0[\;\vp\;]  = \om^2 \vp^2 \; ,
\ee
in which $\omega$ is a control parameter, we define
\be
\label{60}
H_\ep[\;\vp\;]  = H_0[\;\vp\;] + \ep \Dlt H \qquad ( \ep \ra 1 ) \; ,
\ee
where the perturbation term is
\be
\label{61}
 \Dlt H = H - H_0 = ( 1 - \om^2 ) \vp^2 + g\vp^4 \;  .
\ee
Employing perturbation theory with respect to the dummy parameter $\ep$, and
setting $\ep = 1$, leads to the sequence of the approximants
\be
\label{62}
F_k(g,\om) = - \ln \; Z_k(g,\om) \; .
\ee

Control functions for the approximations of odd orders are found from the minimal 
derivative condition
\be
\label{63}
\frac{\prt F_k(g,\om_k)}{\prt \om_k} = 0 \qquad ( k = 1,3,\ldots ) \; .
\ee
For even orders, the above equation does not possess real-valued solutions, 
because of which we set
\be
\label{64}
\om_k = \om_{k-1}(g) \qquad  ( k = 2,4,\ldots ) \;  .
\ee

Thus we obtain the optimized approximants
\be
\label{65}
\overline f_k(g) = F_k(g,\om_k(g) ) \;  .
\ee
Their accuracy can be characterized by the maximal percentage error
\be
\label{66}
 \ep_k = 
\sup_g \left| \; \frac{\overline f_k(g)-f(g)}{f(g)} \;\right|\times 100\% \;  ,
\ee
comparing the optimized approximants with the exact expression (\ref{56}). These 
maximal errors are
$$
\ep_1 = 7\%, \qquad  \ep_2 = 4\%, \qquad \ep_3 = 0.2\%, \qquad 
\ep_4 = 0.2\% \; .
$$
As we see, with just a few terms, we get quite good accuracy, while the bare 
perturbation theory in powers of the coupling parameter $g$ is divergent. Details 
can be found in review \cite{Yukalov_25}.   

This simple model allows for explicitly studying the convergence of the sequence 
of the optimized approximants. It has been proved \cite{Buckley_38,Bender_39} that 
this sequence converges for both ways of defining control functions, either from 
the minimal derivative or minimal-difference conditions.

\section{Eigenvalue Problem}

Another typical example is the calculation of the eigenvalues of Schr\"odinger 
operators, defined by the eigenproblem
\be
\label{67}
 H \psi_n = E_n \psi_n \;  .
\ee
Let us consider a one-dimensional anharmonic oscillator with the Hamiltonian
\be
\label{68}
 H = - \;\frac{1}{2} \; \frac{d^2}{dx^2} + \frac{1}{2}\;x^2 + gx^4 \; ,
\ee
in which $x \in (-\infty, \infty)$ and $g > 0$.

For the initial approximation, we take the harmonic oscillator model 
\be
\label{69}
 H_0 = - \;\frac{1}{2} \; \frac{d^2}{dx^2} + \frac{\om^2}{2}\;x^2    
\ee
with a control parameter $\omega$. Following the approach, we define 
\be
\label{70}
H_\ep = H_0 + \ep \Dlt H \qquad ( \ep \ra 1) \; ,
\ee
where
\be
\label{71}
  \Dlt H = H - H_0 =  \frac{1-\om^2}{2}\;x^2 + g x^4 \;  .
\ee

Employing the Rayleigh-Schr\"{o}dinger perturbation theory with respect to 
the dummy parameter $\ep$, we obtain the spectrum $E_{kn}(g,\om)$, where $k$ 
enumerates the approximation order and $n=0,1,2,\ldots$ is the quantum number 
labeling the states. The zero-order eigenvalue is
\be
\label{72}
E_{0n}(g,\om) = \left( n + \frac{1}{2}\right) \om  \; .
\ee
For odd orders, control functions can be found from the optimization condition
\be
\label{73}
 \frac{\prt}{\prt\om_k} \; E_{kn}(g,\om_k) = 0 \qquad ( k = 1,3,\ldots) \; .
\ee
For even orders, the above equation does not possess real-valued solutions, 
because of which we set
\be
\label{74}
\om_k = \om_{k-1}(g) \qquad ( k = 2,4,\ldots) \; .
\ee

Using optimized perturbation theory results in the eigenvalues 
\be
\label{75}
\overline E_{kn}(g) = E_{kn}(g,\om_k(g))  \;  .
\ee
Comparing these with the numerically found eigenvalues $E_n(g)$ \cite{Hioe_40}, 
we define the percentage errors
\be
\label{76}
 \ep_{kn}(g) = \left|\; 
\frac{\overline E_{kn}(g)-E_n(g)}{E_n(g)} \;\right|  \times 100\% \;  .
\ee
Then we can find the maximal error of the $k$-th order approximation
\be
\label{77}
 \ep_k = \sup_{n,g} \ep_{kn}(g) \; ,
\ee
which gives
$$
\ep_1 = 2\%, \qquad  \ep_2 = 0.8\%, \qquad \ep_3 = 0.8\%, \qquad 
\ep_4 = 0.5\% \;    .
$$
The maximal errors 
\be
\label{78}
 \ep_k^0 \equiv \sup_g \ep_{k0}(g)
\ee
for the ground state are
$$
\ep_1^0 = 2\%, \qquad  \ep_2^0 = 0.8\%, \qquad \ep_3^0 = 0.04\%, \qquad 
\ep_4^0 = 0.03\% \;    .
$$

Again we observe good accuracy and numerical convergence. Recall that the bare 
perturbation theory in powers of the anharmonicity parameter $g$ diverges for any 
finite $g$. The convergence of the sequence of the optimized approximants can be 
proved analytically \cite{Duncan_41,Guida_42}. More details can be found in 
Ref. \cite{Yukalov_25}.

\section{Nonlinear Schr\"{o}dinger Equation}

The method can be applied to strongly nonlinear systems. Let us illustrate this 
by considering the eigenvalue problem 
\be
\label{79}
 H[\;\psi\;] \psi(\br) = E \psi(\br) \;  ,
\ee
with the nonlinear Hamiltonian
\be
\label{80}
H[\;\psi\;]  = -\; \frac{\nabla^2}{2m} + U(\br) + N \Phi_0|\; \psi\;|^2 \; .
\ee
Here $N$ is the number of trapped atoms, the potential
\be
\label{81}
U(\br) = \frac{m}{2} \; \om_\perp^2 \left( x^2 + y^2 +\al^2 z^2 \right)
\ee
is an external potential trapping atoms whose interactions are measured by 
the parameter
\be
\label{82}
 \Phi_0 =  4\pi \; \frac{a_s}{m} \;  ,
\ee
where $a_s$ is a scattering length. This problem is typical for trapped atoms 
in Bose-Einstein condensed state 
\cite{Bogolubov_43,Bogolubov_44,Lieb_45,Letokhov_46,Pethick_47,Yukalov_48,
Bogolubov_49,Yukalov_50}. 

The trap anisotropy is characterized by the trap aspect ratio
\be
\label{83}
 \al \equiv 
\frac{\om_z}{\om_\perp} = \left( \frac{l_\perp}{l_z}\right)^2
\qquad  
\left( l_\perp \equiv \frac{1}{\sqrt{m\om_\perp}}\; , ~ 
l_z \equiv \frac{1}{\sqrt{m\om_z}} \right)    \; .
\ee
It is convenient to introduce the dimensionless coupling parameter
\be
\label{84}
g \equiv 4\pi\; \frac{a_s}{l_\perp} \; N \;  .
\ee
Measuring energy in units of $\om_\perp$ and lengths in units of $l_\perp$, 
we can pass to dimensionless units and write the nonlinear Hamiltonian as
\be
\label{85}
H[\;\psi\;]  = -\; \frac{\nabla^2}{2} + 
\frac{1}{2} \left( r^2 + \al^2 z^2 \right) + g |\; \psi\; |^2  \; ,
\ee
with a dimensionless wave function $\psi$.
 
Applying optimized perturbation theory for the nonlinear Hamiltonian 
\cite{Courteille_51,Yukalov_52}, we take for the initial approximation the 
oscillator Hamiltonian
\be
\label{86}
H_0[\;\psi\;]  = -\; \frac{\nabla^2}{2} + 
\frac{1}{2} \left( u^2 r^2 + v^2 z^2 \right) \;  ,
\ee
in which $u$ and $v$ are control parameters. The zero-order spectrum is given 
by the expression
\be
\label{87}
 E^{(0)}_{nmj} = ( 2n + |\;m\;| + 1 ) u + 
\left( \frac{1}{2} + j \right) v \;  ,
\ee
with the radial quantum number $n=0,1,2,\ldots$, azimuthal quantum number 
$m =0,\pm 1,\pm 2,\ldots$, and the axial quantum number $j=0,1,2,\ldots$. The 
related wave functions are the Laguerre-Hermite modes. The system Hamiltonian 
takes the form
\be
\label{88}
H_\ep =  H_0[\;\psi\;]  + \ep \Dlt H \qquad (\ep \ra 1) \;  ,
\ee
where the perturbation term is 
\be
\label{89}
 \Dlt H \equiv H[\;\psi\;] - H_0[\;\psi\;] = 
\frac{1}{2} \left(1 - u^2 \right) r^2 +
\frac{1}{2}\;\left( \al^2 - v^2 \right) z^2 + g |\;\psi\;|^2 \;  .
\ee

Perturbation theory with respect to the dummy parameter $\varepsilon$ gives the 
energy levels $E^{(k)}_{nmj}$. The control functions are defined by the optimization 
conditions
\be
\label{90}
\frac{\prt}{\prt u_k} \; E_{nmj}^{(k)}(g,u_k,v_k) = 0 \; , \qquad   
\frac{\prt}{\prt v_k} \; E_{nmj}^{(k)}(g,u_k,v_k) = 0
\ee
yielding $u_k = u_k(g)$ and $v_k = v_k(g)$. Applications to trapped atoms are 
discussed in Refs. \cite{Courteille_51,Yukalov_52}.

\section{Hamiltonian Envelopes}

When choosing for the initial approximation a Hamiltonian, one confronts the 
problem of combining two conditions often contradicting each other. From one 
side, the initial approximation has to possess the properties imitating the 
studied problem. From the other side, it has to be exactly solvable, providing 
tools for the explicit calculation of the terms of perturbation theory. If the 
studied Hamiltonian and the Hamiltonian of the initial approximation are too 
much different, perturbation theory, even being optimized, may be poorly 
convergent. In such a case, it is possible to invoke the method of Hamiltonian 
envelopes \cite{Yukalov_25,Yukalov_53}.

\subsection{General Idea}

Suppose we take as an initial approximation a Hamiltonian $H_0$ that, however, 
is very different from the considered Hamiltonian $H$. The difficulty is that 
the set of exactly solvable problems is very limited, so that sometimes it is 
impossible to find another Hamiltonian that would be close to the studied form 
$H$ and at the same time solvable. In that case, we can proceed as follows. 
Notice that, if a Hamiltonian $H_0$ defines the eigenproblem
\be
\label{91}
 H_0 \psi_n = E_n \psi_n \;  ,
\ee
then a function $h(H_0)$ satisfies the eigenproblem
\be
\label{92}
h(H_0) \psi_n = h(E_n) \psi_n   
\ee
enjoying the same eigenfunctions. The function $h(H)$ can be called the 
{\it Hamiltonian envelope} \cite{Yukalov_25,Yukalov_53}. Note that, because of the
property (\ref{92}), $h(H_0)$ can be any real function.

Accepting $h(H_0)$ as an initial Hamiltonian, we obtain the system Hamiltonian
\be
\label{93}
 H_\ep = h(H_0) + \ep \Dlt H \qquad ( \ep \ra 1 ) \;  ,
\ee
with the perturbation term
\be
\label{94}
\Dlt H = H - h(H_0) \;   .
\ee
If we find a function $h(H_0)$ that better imitates the studied system than 
the bare $H_0$, then the convergence of the sequence of approximations can be 
improved.

The general idea in looking for the function $h(H_0)$ is as follows. Let the 
system Hamiltonian be 
\be
\label{95}
H = - \;\frac{\nabla^2}{2m} + V(\br) \;   .
\ee
And let the eigenproblem for a Hamiltonian 
\be
\label{96}
H_0 = - \;\frac{\nabla^2}{2m} + V_0(\br) 
\ee
enjoys exact solutions, although poorly approximating the given system. 
 
Looking for the function $h(H_0)$, we keep in mind that the most influence 
on the behavior of wave functions is produced by the region, where the system 
potential $V(\br)$ displays singular behavior tending to $\pm\infty$. Suppose 
this happens at the point $\br_s$. Then the function $h(H_0)$ has to be chosen 
such that
\be
\label{97}
0 < \lim_{\br\ra\br_s}\; \frac{h(V_0(\br))}{V(\br)} < \infty \;  ,
\ee
that is the function $h(H_0)$ needs to possess the same type of singularity 
as the potential of the studied system. Below we illustrate how this choice 
is done for concrete examples.

\subsection{Power-Law Potentials}

Let us consider the Hamiltonian with a power-law potential
\be
\label{98}
 H = -\; \frac{1}{2m} \; \frac{d^2}{dx^2} + \frac{m\om_0^2}{2} \; x^2 +
A x^\nu \qquad ( \nu > 0 ) \;  ,
\ee
in which $x\in(-\infty,\infty)$, $\om_0>0$, $A>0$, and $\nu>0$. To pass to 
dimensionless units, we scale the energy and length quantities as
\be
\label{99}
 \overline H = \frac{H}{\om_0} \; , \qquad 
\overline x = \sqrt{m\om_0}\; x \;  . 
\ee
The dimensionless coupling parameter is
\be
\label{100}
g \equiv \frac{A}{\om_0}\; ( m \om_0 )^{\nu/2} \;   .
\ee
In what follows, in order not to complicate notation, we omit the bars above 
dimensionless quantities. In dimensionless units, we get the Hamiltonian
\be
\label{101}
 H = -\; \frac{1}{2} \; \frac{d^2}{dx^2} + \frac{x^2}{2} + g x^\nu \; .
\ee
In order to return to the dimensional form, it is sufficient to make the 
substitution 
$$
H \longmapsto \frac{H}{\om_0} \; , \qquad x \longmapsto \sqrt{m\om_0}\; x \; .
$$

Taking for $H_0$ the Hamiltonian
\be
\label{102}
H_0 = -\; \frac{1}{2} \; \frac{d^2}{dx^2} + \frac{u^2}{2} \; x^2 \;  ,
\ee
we compare the potentials
\be
\label{103}
V(x) = \frac{x^2}{2} + g x^\nu \; , \qquad 
V_0(x) = \frac{u^2}{2} \; x^2 \;   .
\ee
As is evident, the singular point here is $x_s=\infty$. To satisfy condition 
(\ref{97}) for $\nu<2$, we have to take
\be
\label{104}
 h(V_0) = V_0 \qquad ( 0 < \nu < 2) \;  ,
\ee
since
$$
 \lim_{x\ra\infty} \; \frac{h(V_0(x))}{V(x)} = u^2 \qquad (\nu < 2) \; ,
$$
while for $\nu>2$, we need to accept 
\be
\label{105}
 h(V_0) = V_0^{\nu/2} \qquad ( \nu > 2) \;  ,
\ee
since now
$$
\lim_{x\ra\infty} \; \frac{h(V_0(x))}{V(x)} = 
\frac{1}{g} \left( \frac{u^2}{2}\right)^{\nu/2} \qquad ( \nu > 2) \; .
$$
In that way, the Hamiltonian envelope is given by the function
\begin{eqnarray}
\label{106}
h(H_0) = \left\{ \begin{array}{lr}
H_0 \; , ~ & ~ 0 < \nu \leq 2 \\
H_0^{\nu/2} \; , ~ & ~ \nu \geq 2 
\end{array} .
\right.  
\end{eqnarray}

\subsection{Inverse Power-Law Potentials} 

The radial Hamiltonian with an inverse power-law potential has the form
\be
\label{107}
H = -\; \frac{1}{2m} \; \frac{d^2}{dr^2} \;  + \; 
\frac{l(l+1)}{2mr^2} \; - \; \frac{A}{r^\nu} \;  ,
\ee
in which ${\bf r} \geq 0$, $l = 0,1,2,\ldots$, $A > 0$, and $\nu > 0$. Again 
we can introduce the dimensionless quantities
\be
\label{108}
 \overline H \equiv \frac{H}{\om} \; , \qquad   
\overline r \equiv \sqrt{m\om} \; r \; ,
\ee
and the dimensionless coupling parameter
\be
\label{109}
 g \equiv \frac{A}{\om} \; (m\om)^{\nu/2} \;  ,
\ee
where $\omega$ is arbitrary. Since $\omega$ is arbitrary, it can be chosen 
such that the coupling parameter be unity,
\be
\label{110}
 g = 1 \; , \qquad \om^{2-\nu} = m^\nu A^2 \;  .
\ee

In dimensionless units the Hamiltonian becomes
\be
\label{111}
H = -\; \frac{1}{2} \; \frac{d^2}{dr^2} \;  + \; 
\frac{l(l+1)}{2r^2} \; - \; \frac{1}{r^\nu} \;   .
\ee
This reminds us the Coulomb problem with the Hamiltonian
\be
\label{112}
H_0 = -\; \frac{1}{2} \; \frac{d^2}{dr^2} \;  + \; 
\frac{l(l+1)}{2r^2} \; - \; \frac{u}{r} \;     .
\ee
Here $u$ is a control parameter. Comparing the potentials
\be
\label{113}
V(r) = -\; \frac{1}{r^\nu} \; , \qquad V_0(r) = -\; \frac{u}{r} \;  ,
\ee
we see that to satisfy condition (\ref{97}) we have to take the envelope 
function as
\be
\label{114}
 h(V_0) = - |\; V_0\;|^\nu \;  ,
\ee
as far as
$$
\frac{h(V_0(r))}{V(r)} = u^\nu \;   .
$$
Then the Hamiltonian envelope reads as
\be
\label{115}
  h(H_0) = - |\; H_0\; |^\nu \qquad ( \nu > 0 ) \; .
\ee

\subsection{Logarithmic Potential}

As one more example, let us take the radial Hamiltonian of arbitrary 
dimensionality with the logarithmic potential
\be
\label{116}
H = -\; \frac{1}{2m} \; \frac{d^2}{dr^2} + \frac{l_d(l_d+1)}{2mr^2} + 
B\ln \; \frac{r}{b}\;  ,
\ee
where $r>0$, $B>0$, $b>0$, and the effective radial quantum number is
\be
\label{117}
l_d \equiv  l + \frac{d-3}{2} \; .
\ee

Again, we need to work with dimensionless quantities, defining
\be
\label{118}
\overline H = mb^2 H \; , \qquad \overline r = \frac{r}{b} \; ,
\ee
and the dimensionless coupling parameter
\be
\label{119}
g \equiv m b^2 B \;   .
\ee
Then, for the simplicity of notation, we omit the bars over the letters and 
get the dimensionless Hamiltonian
\be
\label{120}
H = -\; \frac{1}{2} \; \frac{d^2}{dr^2} + \frac{l_d(l_d+1)}{2r^2} + 
 g \ln \; r \;   .
\ee
Accepting at the starting step the oscillator Hamiltonian
\be
\label{121}
H_0 = -\; \frac{1}{2} \; \frac{d^2}{dr^2} + \frac{l_d(l_d+1)}{2r^2} + 
\frac{u^2}{2} \; r^2  \; ,
\ee
we have to compare the potentials
\be
\label{122}
 V(r) = g \ln \; r \; , \qquad V_0(r) =\frac{u^2}{2} \; r^2 \;  .
\ee
Now the singular points are $r_s = 0$ and $r_s = \infty$. This dictates the 
choice of the envelope function
\be
\label{123}
h(V_0) = \ln\; V_0  \;   ,
\ee
since
$$
\lim_{r\ra 0} \; \frac{h(V_0(r))}{V(r)} =  
\lim_{r\ra\infty} \; \frac{h(V_0(r))}{V(r)} = \frac{2}{g} \; .
$$
Some explicit calculations can be found in Refs. \cite{Yukalov_25,Yukalov_53}.

Optimized perturbation theory, whose main points are expounded above, has been 
applied to a great variety of problems in statistical physics, condensed matter 
physics, chemical physics, quantum field theory, etc, as is reviewed in 
Ref. \cite{Yukalov_25}.

\section{Optimized Expansions: Summary}

As is explained above, the main idea of optimized perturbation theory is the 
introduction of control parameters that generate order-dependent control functions 
controlling the convergence of the sequence of optimized approximants. Control 
functions can be incorporated in the perturbation theory in three main ways: by 
choosing an initial approximation containing control parameters, by making a change
of variables and resorting to a reexpansion trick, or by accomplishing a 
transformation of the given perturbation sequence. Control functions are defined
by optimization conditions. Of course, there are different variants of implanting
control functions and choosing the appropriate variables. In some cases, control 
functions $u_k(x)$ can become control parameters $u_k$, since constants are just a 
particular example of functions. 

Below we summarize the main ideas shedding light on the common points for choosing 
control functions, the variables for expansions, on the convergence of the sequence 
of optimized approximants, and on the examples when control functions can be reduced
to control parameters. Also, we shall compare several methods of optimization. To 
make the discussion transparent, we shall illustrate the ideas on the example of a 
partition function for a zero-dimensional $\varphi^4$ field theory and on the model 
of one-dimensional anharmonic oscillator.

\subsection{Expansion over Dummy Parameters}

The standard and often used scheme of optimized perturbation theory is based on the 
incorporation of control functions through initial approximations, as is mentioned 
in Sec. 4.1. Suppose we deal with a Hamiltonian $H(g)$ containing a physical
parameter $g$, say coupling parameter. When the problem cannot be solved exactly,
one takes a trial Hamiltomian $H_0(u)$ containing control parameters denoted through 
$u$. One introduces the Hamiltonian 
\be
\label{A1}
H_\ep(g,u) = H_0(u) +\ep [ \; H(g) - H_0(u) \; ] \;   ,
\ee 
in which $\varepsilon$ is a dummy parameter. One calculates the quantity of interest
$F_k(g,u,\varepsilon)$ by means of perturbation theory in powers of the dummy parameter
$\varepsilon$,
\be
\label{A2}
F_k(g,u,\ep) = \sum_{n=0}^k \; c_n(g,u) \; \ep^n \; ,
\ee
after which sends this parameter to one, $\varepsilon\ra 1$. 

Employing one of the optimization conditions discussed in Sec. 6, one finds the 
control functions $u_k(g)$. The most often used optimization conditions are the 
minimal-difference condition
\be
\label{A3}
F_k(g,u,1) - F_{k-1}(g,u,1) = 0 \; , \qquad  u = u_k(g)
\ee
and the minimal-derivative condition
\be
\label{A4}
 \frac{\prt}{\prt u} \; F_k(g,u,1) = 0 \; , \qquad u = u_k(g) \;  .
\ee
Substituting the found control functions $u_k(g)$ into $F_k(g,u_k(g),1)$ results 
in the optimized approximants
\be
\label{A5}
\overline F_k(g) = F_k(g,u_k(g),1) \;  .
\ee
This scheme of optimized perturbation theory was suggested and employed in Refs. 
\cite{Yukalov_16,Yukalov_17,Yukalov_18,Yukalov_19,Yukalov_20,Yukalov_21} and in 
numerous following publications, as can be inferred from the review works
\cite{Yukalov_24,Yukalov_25,Dineykhan_26,Sissakian_27,Feranchuk_28}. As is evident, 
the same scheme can be used dealing with Lagrangians or action functionals.     
 
Instead of the notation $\varepsilon$ for the dummy parameter, it is admissible to 
use any other letter, which, as is clear, is of no importance. Sometimes one denotes 
the dummy parameter as $\delta$ and, using the same standard scheme, one calls it 
delta expansion. However, using a different notation does not compose a different 
method.

\subsection{Scaling Relations: Partition Function}

The choice of variables for each particular problem is the matter of convenience. 
Often it is convenient to use the combinations of parameters naturally occurring
in the considered case. These combinations can be found from the scaling relations 
available for the considered problem. 

Let us start with the simple, but instructive, case of the integral representing 
the partition function (or generating functional) of the so-called zero-dimensional 
$\varphi^4$ field theory
\be
\label{A6}
 Z(g,\om_0) = 
\frac{1}{\sqrt{\pi}} \; \int_{-\infty}^\infty \exp( - H[\;\vp\;] ) \; d\vp \;  ,
\ee
with the Hamiltonian
\be
\label{A7}  
H[\;\vp\;] = \om_0^2 \vp^2 + g\vp^4 \; ,
\ee
where $g > 0$. 

Invoking the scaling $\varphi\longmapsto \lambda \varphi$ leads to the relation
\be
\label{A8}
 Z(g,\om_0) = \lbd Z\left( \lbd^4 g , \lbd^2\om_0 \right) \;  .
\ee
By  setting $\lambda = g^{-1/4}$ yields the equality
\be
\label{A9}
  Z(g,\om_0) =  
\frac{1}{g^{1/4}} \; Z\left(1,\frac{\om_0}{\sqrt{g}} \right) \;  .
\ee
And setting $\lambda = \omega_0^{-1/2}$ gives
\be
\label{A10}
  Z(g,\om_0) = \frac{1}{\sqrt{\om_0} } \; 
Z\left( \frac{g}{\om_0^2}\;, 1 \right) \; .
\ee
These relations show that at large coupling constant the expansion is realized 
over the combination $\omega_0/\sqrt{g}$, while at small coupling constant the 
natural expansion is over $g/\omega_0^2$.

\subsection{Scaling Relations: Anharmonic Oscillator}

The other typical example frequently treated for demonstrational purposes is the 
one-dimensional anharmonic oscillator with the Hamiltonian
\be
\label{A11}
H = - \; \frac{1}{2}\; \frac{\prt^2}{\prt x^2} + \frac{\om_0^2}{2}\; x^2 +
g x^p \qquad (p > 0 ) \;   ,
\ee
where $g > 0$. Let the energy levels $E(g,\omega_0)$ of the Hamiltonian be of 
interest. 

By scaling the spatial variable $x\longmapsto \lambda x$ results in the relation
\be
\label{A12}
E(g,\om_0) = \lbd^{-2} \; 
E\left( \lbd^{p+2} g, \lbd^2 \om_0 \right) \;   .
\ee
Setting $\lambda = g^{-1/(1+p/2)}$ gives
\be
\label{A13}
E(g,\om_0) = g^{1/(1+p/2)} \; E\left( 1, \frac{\om_0}{g^{1/(1+p/2)}} \right) \; ,
\ee
while for $\lambda = \omega_0^{-1/2}$ we get the relation
\be
\label{A14}
 E(g,\om_0) = \om_0  E\left( \frac{g}{\om_0^{1+p/2}}\;, 1 \right) \; .
\ee
In particular, for the quartic anharmonic oscillator, with $p=4$, we have
\be
\label{A15}
 E(g,\om_0) = g^{1/3} \; E\left( 1, \frac{\om_0}{g^{1/3}} \right)
\ee
and 
\be
\label{A16}
 E(g,\om_0) = 
\om_0 \; E\left( \frac{g}{\om_0^3}\;, 1 \right) \qquad ( p = 4) \; .
\ee
Again these relations suggest what are the natural variables for expansions over 
large or small coupling constants.

\subsection{Optimized Expansion: Partition Function}

The standard scheme of the optimized perturbation theory has been applied to the 
model (\ref{A6}) many times, accepting as an initial Hamiltonian the form
\be
\label{A17}
H_0 = \om^2 \vp^2 \;   ,
\ee
in which $\omega$ is a control parameter. Then Hamiltonian (\ref{A1}) becomes
\be
\label{A18}
H_\ep = \om^2\vp^2 + 
\ep \left[\; ( \om_0^2 - \om^2 ) \vp^2 + g \vp^4 \; \right] \; .
\ee
Note that Hamiltonian (\ref{A7}) transforms into (\ref{A18}) by means of the 
replacement
\be
\label{A19}
 \om_0^2 \longmapsto \om^2 + \ep (\om_0^2 - \om^2 ) \; , \qquad
g \longmapsto \ep g \;  .
\ee
Following the standard scheme of optimized perturbation theory for the partition 
function, and using the optimization conditions for defining control functions, it 
was found \cite{Buckley_38,Bender_39,Guida_42} that at large orders the control 
functions behave as
\be
\label{A20}
 \om_k(g) \simeq \al \om_0 (gk)^{1/4} \qquad ( k \ra \infty) \;  .
\ee
The minimal-difference and minimal-derivative conditions give $\alpha=1.0729855$. 
It was proved \cite{Buckley_38,Bender_39} that this scheme results in the sequence 
of optimized approximants for the partition function that converges to the exact 
numerical value. The convergence occurs for any $\alpha>\alpha_c = 0.9727803$.

\subsection{Optimized Expansion: Anharmonic Oscillator}

The one-dimensional quartic anharmonic oscillator with the Hamiltonian (\ref{A11}), 
where $p=4$ and $g > 0$, also serves as a typical touchstone for testing approximation 
methods. The initial approximation is characterized by the harmonic oscillator
\be
\label{A21}
 H_0 = -\; \frac{1}{2}\;\frac{\prt^2}{\prt x^2} + \frac{\om^2}{2} \; x^2 \; ,
\ee
in which $\omega$ is a control parameter. The Hamiltonian (\ref{A1}) takes the 
form
\be
\label{A22}
H_\ep  =  -\; \frac{1}{2}\;\frac{\prt^2}{\prt x^2} + \frac{\om^2}{2} \; x^2
+ \ep \left[\; \frac{1}{2} \; ( \om_0^2 - \om^2 ) x^2 + g x^4 \;\right] \;  .
\ee
As is seen, the transformation from (\ref{A11}) to (\ref{A22}) is realized by 
the same substitution (\ref{A19}), with the substitution for $\omega_0$ that can 
be represented as
\be
\label{A23}
 \om_0^2 \longmapsto \om^2 \left[\; 1 \; - 
\ep \left( 1 - \; \frac{\om_0^2}{\om^2} \right) \; \right] \;  .
\ee 
This shows the appearance of the characteristic combination $1-(\om_0/\om)^2$ that
will be used below.  

Calculating the energy eigenvalues following the standard scheme, one finds 
\cite{Duncan_41,Guida_42} the control function
\be
\label{A24}  
\om_k(g) \simeq \al \om_0(gk)^{1/3} \qquad ( k \ra \infty) \;  ,
\ee
with $\al\approx 1$ for both the minimal-difference and minimal-derivative 
conditions. The convergence of the sequence of optimized approximants to the exact 
numerical values \cite{Hioe_40}, found from the solution of the Schr\"{o}dinger 
equation, takes place for $\al>\al_c=0.9062077$.

\section{Order-Dependent Mapping}

Sometimes the procedure can be simplified by transforming the initial expansion, 
say in powers of a coupling constant, into expansions in powers of other parameters. 
By choosing the appropriate change of variables, it can be possible to reduce the 
problem to the form where control functions $u_k(g)$ are downgraded to control 
parameters $u_k$. The change of variables depends on the approximation order, 
because of which it is called the order-dependent mapping \cite{Seznec_54}.

\subsection{Change of Variables}

Let us be given an expansion in powers of a variable $g$,
\be
\label{A25}
 f_k(g) = \sum_{n=0}^k a_n g^n \;  .
\ee
By analyzing the properties of the considered problem, such as its scaling 
relations and the typical combinations of parameters arising in the process of 
deriving perturbative series, it is possible to notice that it is convenient to 
denote some parameter combinations as new variables. Then one introduces the 
change of variables
\be
\label{A26}
g = g(z,u) = u y(z) \; ,
\ee
where
\be
\label{A27}
u = \frac{g}{y(z)} = u(g,z)   
\ee
is treated as a control parameter. By substituting (\ref{A26}) into (\ref{A25}) 
gives the function $f_k(g(z,u))$, which has to be expanded in powers of $z$ up to 
order $k$, leading to the series
\be
\label{A28} 
F_k(z,u) = \sum_{n=0}^k b_n(u) z^n \;  .
\ee
The minimal-difference condition 
\be
\label{A29}
F_k(z,u) - F_{k-1}(z,u) = 0
\ee
yields the equation 
\be
\label{A30}
b_k(u) = 0 \; , \qquad u = u_k
\ee
defining the control parameters $u_k$. Since, according to (\ref{A27}), the value 
$u_k$ denotes the combination of parameters $u_k = u_k(g,z)$, hence it determines 
the control functions $z_k(g)$. The pair $u_k$ and $z_k(g)$, being substituted into 
(\ref{A28}), results in the optimized approximants
\be
\label{A31}
\overline F_k(g) = F_k(z_k(g),u_k) \;  .
\ee
Thus, the convenience of the chosen change of variables is in the possibility of 
dealing at the intermediate step with control parameters instead of control functions 
that appear at a later stage.

\subsection{Partition Function}

To illustrate the method, let us consider the partition function (\ref{A6}) following
the described scheme \cite{Seznec_54}. From the substitution (\ref{A19}) it is clear 
that natural combinations of parameters appearing in perturbation theory with respect 
to the term with $\varepsilon$ in the Hamiltonian (\ref{A18}) are 
\be
\label{A32}
 z = \frac{\om^2- \om_0^2}{\om^2} = 1 \; - \; \frac{\om_0^2}{\om^2}  
\ee
and 
\be
\label{A33}
y(z) = \frac{\om^2(\om^2-\om_0^2)}{\om_0^4} = \frac{z}{(1-z)^2} \; .
\ee
Then the combination of parameters (\ref{A27}) reads as
\be
\label{A34}
u = \frac{g}{z} \; ( 1 - z)^2 = \frac{g\om_0^4}{\om^2(\om^2-\om_0^2)}
\ee

In order to simplify the notation, it is possible to notice that the parameter 
$\om$ always enters the equations being divided by $\omega_0$. Therefore, measuring 
$\om$ in units of $\omega_0$ is equivalent to setting $\om_0\ra 1$. In these units,
$$
 z = 1 \; - \; \frac{1}{\om^2} \; , \qquad u = \frac{g}{\om^2(\om^2-1)} \;  .
$$
Finding from the minimal-difference condition (\ref{A29}) the control parameter $u_k$
and using definition (\ref{A34}) gives the control function
\be
\label{A35}
 z_k(g) = 1\; - \; \frac{\sqrt{u_k^2+4g u_k}\; - u_k}{2g} \;  . 
\ee
Then relation (\ref{A32}) results in the control function
\be
\label{A36}
 \om_k(g) = \frac{1}{\sqrt{1-z_k(g)}} = \frac{1}{\sqrt{2}} \;
\left( 1 + \sqrt{1 + \frac{4g}{u_k} } \right)^{1/2} \;  .
\ee
Finally, one gets the partition function $Z_k(z_k(g),u_k)$. 

This procedure, with the change of variables used above, has been shown 
\cite{Guida_120} to be equivalent to the standard scheme of optimized perturbation 
theory resulting in optimized approximants $\overline{Z}_k(g)$.

\subsection{Anharmonic Oscillator}

Again using the dimensionless units, as in the previous section, one sets the 
notations
\be
\label{A37}
 y(z) = \frac{z}{(1-z)^{3/2} } \; , \qquad
 z = 1 \; - \; \frac{1}{\om^2} \;  .
\ee
Then the combination (\ref{A27}) becomes
\be
\label{A38}
u = \frac{g}{\om(\om^2-1)} = \frac{g}{z}\; ( 1 - z)^{3/2} \;  .
\ee
Similarly to the previous section, one finds the control parameter $u_k$ and from 
(\ref{A38}) one obtains the control functions $z_k(g)$ and $\om_k(g)$. The resulting
energy levels $E_k(z_k(g),u_k)$ coincide with the optimized approximants 
$\overline{E}_k(g)$, as has been proved in \cite{Guida_120}.

\section{Variational Expansions}

The given expansion over the coupling constant (\ref{A25}) can be reexpanded with 
respect to other variables in several ways. One of the possible reexpansions has been 
termed variational perturbation theory \cite{Kleinert_29}. Below it is illustrated by 
the example of the anharmonic oscillator in order to compare this type of a reexpansion 
with other methods. 

Let us consider the energy levels of the anharmonic oscillator with the Hamiltonian 
(\ref{A11}) with $p=4$. As is clear from the scaling relations of Sec. 11, the energy 
can be represented as an expansion 
\be
\label{A39}
E_k(g,\om_0) = 
\om_0 \sum_{n=0}^k c_n\left( \frac{g}{\om_0^3}\right)^n \; .
\ee

We have the identity
\be
\label{A40}
\om_0^2 = \om^2 + \om_0^2 - \om^2
\ee
that is a particular case of the substitution (\ref{A19}) with the control parameter 
$\om$ and $\ep= 1$. Employing the notation
\be
\label{A41}
 z = 1 \; - \; \frac{\om_0^2}{\om^2} = \frac{g\om_0^3}{\om^3 u} \;  ,
\ee
where  
\be
\label{A42}
 u = \frac{g\om_0^3}{\om^3 z} = \frac{g\om_0^3}{\om(\om^2-\om_0^2)} =
\frac{g}{z} \; (1 - z)^{3/2} \; ,
\ee
it is straightforward to rewrite the identity (\ref{A40}) 
in the form
\be
\label{A43}
 \om_0 = \om\; \sqrt{1-z} = \om \; 
\sqrt{1 \; - \; \frac{g}{\om^3 u} } \;  .
\ee
This form is substituted into expansion (\ref{A39}), which then is reexpanded in 
powers of the new variable $g/\omega^3$, while keeping $u$ untouched and setting 
$\omega_0$ to one. The reexpanded series is truncated at order $k$. Comparing this 
step with the expansion in Sec. 11, it is evident that this is equivalent to the 
expansion over the dummy parameter $\varepsilon$. And comparing the expansion over 
$g/\omega^3$ with the expansion over $z$ in Sec. 12, we see that they are also 
equivalent. Thus we come to the expansion
\be
\label{A44}
E_k(g,\om) = \om \sum_{n=0}^k d_n(u) \;\left( \frac{g}{\om^3}\right)^n \;  ,
\ee
where
$$
d_n(u) = \sum_{j=0}^n C_{nj} \; \left( - \; \frac{1}{u} \right)^{n-j} \; .
$$
Then one substitutes back the expression (\ref{A42}) for $u=u(g,\omega)$. 

The control function $\omega_k(g)$ is defined by the minimal derivative condition, 
or, when the latter does not have real solutions, by the zero second derivative over 
$\omega$ of the energy $E_k(g,\omega)$. The found control function $\omega_k(g)$ is
substituted into $E_k(g,\omega)$, thus giving the optimized approximant 
\be
\label{A45}
 \overline E_k(g) = E_k(g,\om_k(g) ) \;  .
\ee

The equivalence of the above expansion in powers of $g/\om^3$ to the expansions with 
respect to the dummy parameter $\ep$, or with respect to the parameter $z$, becomes 
evident if we use the notation of the present section and notice that the substitution 
(\ref{A23}) can be written as
\be
\label{A46}
 \om_0 \longmapsto \om\; \sqrt{1-\ep z} = 
\om\; \sqrt{ 1 -\; \frac{\ep g}{\om^3 u} } \;  .
\ee
This makes it immediately clear that the expansion over $g/\om^3$, with keeping 
$u$ untouched, is identical to the expansion over the dummy parameter $\ep$.

\section{Control Functions and Control Parameters}

It is important to remark that it is necessary to be cautious introducing control 
functions through the change of variables and reexpansion. Strictly speaking, such 
a change cannot be postulated arbitrarily. When the change of variables is analogous 
to the procedure of using the substitutions, such as (\ref{A19}), (\ref{A23}) or 
(\ref{A46}), naturally arising in perturbation theory, as in Sec. 4, then the results
of these variants will be close to each other. However, if the change of variables
is arbitrary, the results can be not merely inaccurate, but even qualitatively 
incorrect \cite{Yukalov_25,Aoyama_121}.

It is also useful to mention that employing the term control functions, we keep 
in  mind that in particular cases they can happen to become parameters, although 
order-dependent. Then instead of functions $u_k(x)$ we can have parameters $u_k$.
There is nothing wrong in this, as far as parameters are a particular example of 
functions. The reduction of control functions to control parameters can occur in 
the following cases. 

It may happen that in the considered problem there exists such a combination of
characteristics that compose the quantities $u_k$ depending only on the approximation
order but not depending on the variable $x$. For instance, this happens in the mapping
of Sec. 12, where the combinations $u_k = u_k(g,\omega_k(g)$ play the role of control 
parameters. In the case of the partition function, this is the combination (\ref{A34})
and for the anharmonic oscillator, it is the combination (\ref{A38}).    

The other example is the existence in the applied optimization of several 
conditions restricting the choice of control parameters. The typical situation is 
when the optimization condition consists in the comparison of asymptotic expansions 
of the sought function and of the approximant. Suppose that, in addition to the 
small-variable expansion
\be
\label{A47}
 f_k(x) = \sum_{n=0}^k a_n x^n \qquad ( x \ra 0 ) \;  ,
\ee
we know the large-variable expansion of the sought function 
\be
\label{A48} 
 f(x) \simeq \sum_{n=0}^p b_n\; \frac{1}{x^n} \qquad ( x \ra \infty) \; .
\ee

Let us assume that we have found the optimized approximant $F_k(x,u_k(x))$, where 
the control functions $u_k(x)$ are defined by one of the optimization conditions of 
Sec. 6. These conditions provide a uniform approximation of the sought function on the 
whole interval of its definition. However the resulting approximants $F_k(x,u_k(x))$
are not required to give exact coefficients of asymptotic expansions either at small
or at large variable $x$. If we wish that these asymptotic coefficients would exactly
coincide with the coefficients of the known asymptotic expansions (\ref{A47}) and
(\ref{A48}), then we have to implant additional control parameters and impose 
additional asymptotic conditions. This can be done by using the method of corrected 
Pad\'{e} approximants \cite{Yukalov_25,Gluzman_84,Gluzman_85,Gluzman_86,Wellenhofer_122}. 
To this end, we define the optimized approximant as
\be
\label{A49}
 \overline f_k(x) = F_k(x,u_k(x) ) P_{N/N}(x) \; ,
\ee
where
\be
\label{A50}
P_{N/N}(x) = \frac{a_0+\sum_{n=1}^N c_n x^n}{1 + \sum_{n=1}^N d_n x^n}
\ee
is a diagonal Pad\'{e} approximant, whose coefficients $c_n$ and $d_n$, playing the 
role of control parameters, are prescribed by the accuracy-through-order procedure, 
so that the asymptotic expansions of (\ref{A49}) would coincide with the given 
asymptotic expansions of the sought function at small $x$,
\be
\label{A51}
 \overline f_k(x) \simeq \sum_{n=0}^k a_n x^n \qquad ( x \ra 0 )
\ee
and at large $x$,
\be
\label{A52}
 \overline f_k(x) \simeq \sum_{n=0}^p b_n \; \frac{1}{x^n} \qquad 
( x \ra \infty ) \; .
\ee
The number of the parameters in the Pad\'{e} approximant is such that to satisfy the 
imposed asymptotic conditions (\ref{A51}) and (\ref{A52}).

\section{Self-Similar Approximation Theory}

As has been emphasized above, the idea of introducing control functions for the 
purpose of governing the convergence of a sequence stems from the optimal 
control theory, where one introduces control functions in order to regulate 
the trajectory of a dynamical system, for instance so that to force the 
trajectory to converge to a desired point. The analogy between perturbation 
theory and the theory of dynamical systems has been strengthened even more in 
the self-similar approximation theory 
\cite{Yukalov_24,Yukalov_25,Yukalov_54,Yukalov_55,Yukalov_56,Yukalov_57,Yukalov_58}. 
The idea of this theory is to consider the transfer from one approximation to 
another as the motion on the manifold of approximants, where the approximation 
order plays the role of discrete time.

Suppose, after implanting control functions, as explained in Sec.4, we have 
the sequence of approximants $F_k(x,u_k)$. Recall that the control functions can 
be defined in different ways, as has been discussed above. Therefore we, actually, 
have the manifold of approximants associated with different control functions, 
\be
\label{124}
 \mathbb{A} = \{ F_k(x,u_k) : ~ 
\mathbb{R}\times\mathbb{R} \longmapsto \mathbb{R} ; ~ k=0,1,2,\ldots \} \; .
\ee
This will be called the {\it approximation manifold}. Generally, it could be 
possible to define a space of approximants. However the term approximation space 
is used in mathematics in a different sense \cite{Pietsch_59}. So, we shall deal 
with the approximation manifold. The transfer from an approximant $F_k$ to another 
approximant $F_{k+p}$ can be understood as the motion with respect to the discrete 
time, whose role is played by the approximation order $k$. The sequence of 
approximants $F_k(x,u_k)$ with a fixed choice of control functions $u_k=u_k(x)$ 
defines a trajectory on the approximation manifold (\ref{124}).

Let us fix the rheonomic constraint
\be
\label{125}
 F_0(x,u_k(x)) = f \; , \qquad x = x_k(f) \; ,  
\ee
defining the expansion function $x_k(f)$. Recall that in the theory of dynamical 
systems a rheonomic constraint is that whose constraint equations explicitly 
contain or are dependent upon time. In our case, time is the approximation order 
$k$. The inverse constraint equation is 
\be
\label{126}
x_k(F_0(x,u_k(x))) = x \;  .
\ee

Let us introduce the endomorphism 
\be
\label{127}
y_k(f) : ~ \mathbb{Z}_+\times\mathbb{R} \longmapsto \mathbb{R} \;  ,
\ee
by the definition acting as
\be
\label{128}
y_k(f) \equiv F_k( x_k(f), u_k( x_k(f) ) ) \;   .
\ee
This endomorphism and the approximants are connected by the equality 
\be
\label{129}
 y_k(F_0(x,u_k(x))) = F_k(x,u_k(x)) \;  .
\ee
The set of endomorphisms forms a dynamical system in discrete time
\be
\label{130}
\{ y_k(f) : ~ \mathbb{Z}_+\times\mathbb{R} \longmapsto \mathbb{R} \} \;  ,
\ee
with the initial condition
\be
\label{131}
 y_0(f) = f \;  .
\ee
By this construction, the sequence of endomorphisms $\{y_k(f)\}$, forming 
the dynamical system trajectory, is bijective to the sequence of approximants 
$\{F_k(x,u_k(x))\}$. Since control functions, by default, make the sequence of 
approximants $F_k(x,u_k(x))$ convergent, this means that there exists a limit
\be
\label{132}
 F^*(x) = \lim_{k\ra\infty} F_k(x,u_k(x)) \;  .
\ee
And as far as the sequence of approximants is bijective to the trajectory of 
the dynamical system, there should exist the limit
\be
\label{133}
 y^*(f) = \lim_{k\ra\infty} y_k(f) \;  .
\ee
This limit, being the final point of the trajectory, implies that it is a fixed 
point, for which 
\be
\label{134}
y_p(y^*(f) ) = y^*(f) \qquad ( p \geq 0 ) \;  .
\ee
Thus to find the limit of an approximation sequence is equivalent to determining 
the fixed point of the dynamical system trajectory. 

We may notice that for large $p$, the self-similar relation holds:
\be
\label{135}
 y_{k+p}(f) \simeq y_k(y_p(f) ) \qquad ( p \ra \infty ) \;  ,
\ee
which follows from conditions (\ref{133}) and (\ref{134}). As far as in the 
real situations it is usually impossible to reach the limit of infinite 
approximation order, we assume the validity of the {\it self-similar relation} 
for finite approximation orders:
\be
\label{136}
  y_{k+p}(f) = y_k(y_p(f) ) \; .
\ee
This relation implies the semi-group property
\be
\label{137}
 y_k \cdot y_p = y_{k+p} \; , \qquad y_0 = 1 \;  .
\ee
The dynamical system in discrete time (\ref{130}) with the above semi-group 
property is called cascade (semicascade). The theory of such dynamical systems 
is well developed \cite{Walker_60,Hale_61}. In our case, this is an {\it 
approximation cascade} \cite{Yukalov_25}.    

Since, as is said above, in realistic situations we are able to deal only with 
finite approximation orders, we can find not an exact fixed point $y^*(f)$, but 
an approximate fixed point $y^*_k(f)$. The corresponding approximate limit of 
the considered sequence is
\be
\label{138}
F^*_k(x,u_k(x)) = y_k^*(F_k(x,u_k(x))) \; .
\ee
If the form $F_k(x,u_k)$ is obtained by means of a transformation 
\be
\label{139}
 F_k(x,u_k) = \hat T[\; u\;]\; f_k(x) \;  ,
\ee
like in (\ref{27}), then the resulting self-similar approximant reads as
\be
\label{140}
 f_k^*(x) = \hat T^{-1}[\; u\;]\; F^*_k(x,u_k(x)) \;  .
\ee

\section{Embedding Cascade into Flow}

Usually, it is more convenient to deal with dynamical systems in continuous 
time than with systems in discrete time. For this purpose, it is possible to 
embed the approximation cascade into an {\it approximation flow}, which is 
denoted as
\be
\label{141}
\{ y_k(f) : ~ \mathbb{Z}_+\times\mathbb{R} \longmapsto \mathbb{R} \} 
\in \{ y(t,f) : ~ \mathbb{R}_+\times\mathbb{R} \longmapsto \mathbb{R} \} 
\ee
and implies that the endomorphism in continuous time enjoys the same group 
property as the endomorphism in discrete time,
\be
\label{142}
 y(t+t',f) = y(t,y(t',f)) \;  ,
\ee
that the flow trajectory passes through all points of the cascade trajectory,
\be
\label{143}
 y(t,f) = y_k(f) \qquad (t = k ) \;  ,
\ee
and starts from the same initial point,
\be
\label{144}
y(0,f) = f \; .
\ee
 
The self-similar relation (\ref{142}) can be represented as the Lie equation
\be
\label{145}
 \frac{\prt}{\prt t} \; y(t,f) = v(y(t,f)) \;  ,
\ee
in which $v(y(t,f))$ is a velocity field. Integrating the latter equation 
yields the evolution integral
\be
\label{146}
 \int_{y_k}^{y_k^*} \frac{dy}{v(y)} = t_k \; ,
\ee
where $t_k$ is the time required for reaching the fixed point $y^*_k=y^*_k(f)$ 
from the approximant $y_k=y_k(f)$. Using relations (\ref{129}) and (\ref{138}), 
this can be rewritten as 
\be
\label{147}
 \int_{F_k}^{F_k^*} \frac{df}{v_k(f)} = t_k \;  ,
\ee
where $F_k=F_k(x,u_k(x))$ and $F^*_k=F^*_k(x,u_k(x))$. 

The velocity field can be represented resorting to the Euler discretization
\be
\label{148}
v_k(f) = y_{k+1}(f) - y_k(f) \;  .
\ee
This is equivalent to the form
\be
\label{149}
 v_k(f) = F_{k+1}(x_{k+1},u_{k+1}) - F_k(x_k,u_k) \;  ,
\ee
in which
$$
x_k =x_k(f) \; , \qquad u_k = u_k(x_k) = u_k(x_k(f)) \; .
$$

We may notice that the velocity field is directly connected with the Cauchy 
difference (\ref{39}), since
\be
\label{150}
 v_k(f) = C(F_{k+1},F_k) \; .
\ee
As is explained in Sec. 6, the Cauchy difference of zero order equals zero, 
hence in that order the velocity is zero, and $F^*_k = F_k$. The Cauchy 
difference of first order is nontrivial, being given by expression (\ref{46}). 
In this order, the velocity field becomes
\be
\label{151}
 v_k(f) = F_{k+1}(x_k,u_k) - F_k(x_k,u_k) + (u_{k+1}-u_k) \;
\frac{\prt}{\prt u_k}\; F_k(x_k,u_k) \;  .
\ee
The smaller the velocity, the faster the fixed point is reached. Therefore 
control functions should be defined so that to make the velocity field minimal:
\be
\label{152}
\min_u |\; v_k(f)\;| = \min_u |\; C(F_{k+1},F_k)\;|  \; .
\ee
Thus we return to the optimization conditions of optimized perturbation theory, 
discussed in Sec. 6. Opting for the optimization condition
\be
\label{153}
(u_{k+1}-u_k) \; \frac{\prt}{\prt u_k}\; F_k(x_k,u_k) = 0
\ee
simplifies the velocity field to the form
\be
\label{154}
  v_k(f) = F_{k+1}(x_k,u_k) - F_k(x_k,u_k) \;  .
\ee

\section{Stability Conditions}

The sequence $\{y_k(f)\}$ defines the trajectory of the approximation cascade 
that is a type of a dynamical system. The motion of dynamical systems can be 
stable or unstable. The stability of motion for the approximation cascade can 
be characterized \cite{Yukalov_25,Yukalov_58,Yukalov_62} similarly to the 
stability of other dynamical systems \cite{Walker_60,Ott_63,Schuster_64}. 
Dealing with real problems, one usually considers finite steps $k$. Therefore 
the motion stability can be defined only locally.

The local stability at the $k$-th step is described by the local map multiplier
\be
\label{155}
 \mu_k(f) \equiv \frac{\dlt y_k(f)}{\dlt y_0(f)} = 
\frac{\prt y_k(f)}{\prt f} \; .
\ee
The motion at the step $k$, starting from an initial point $f$, is stable when
\be
\label{156}
|\; \mu_k(f) \; | < 1 \;   .
\ee
The maximal map multiplier
\be
\label{157}
\mu_k \equiv \sup_f |\; \mu_k(f) \; |
\ee
defines the global stability with respect to $f$, provided that
\be
\label{158}
 \mu_k < 1 \;  .
\ee
The maximum is taken over all admissible values of $f$. 

The image of the map multiplier (\ref{155}) on the manifold of the variable 
$x$ is
\be
\label{159}
m_k(x) = \mu_k(F_0(x,u_k(x) ) \; . 
\ee
The motion at the $k$-th step at the point $x$ is stable if
\be
\label{160}
 |\; m_k(x) \;| < 1 \;  .
\ee
Respectively, the motion is globally stable with respect to the domain of $x$ 
when the maximal map multiplier
\be
\label{161}
 m_k \equiv \sup_x  |\; m_k(x) \;|
\ee
is such that
\be
\label{162}
 m_k < 1 \; .
\ee

The map multiplier at the fixed point $y^*_k(f)$ is
\be
\label{163}
 \mu^*_k(f) \equiv \frac{\prt y^*_k(f)}{\prt f} \;  .
\ee
The fixed point is locally stable when
\be
\label{164}
  |\; \mu^*_k(f) \;| < 1 \;  ,   
\ee
and it is globally stable with respect to $f$ if the maximal multiplier
\be
\label{165}
 \mu^*_k \equiv \sup_f  |\; \mu^*_k(f) \;| \; 
\ee
satisfies the inequality
\be
\label{166}
\mu^*_k < 1 \;  .
\ee

The above conditions of stability can be rewritten in terms of the local 
Lyapunov exponents
\be
\label{167}
\lbd_k(f) \equiv \frac{1}{k} \; \ln |\; \mu_k(f) \;| \; , \qquad 
\lbd^*_k(f) \equiv \frac{1}{k} \; \ln |\; \mu^*_k(f) \;| \;   .
\ee
The motion at the $k$-th step is stable provided the Laypunov exponents are 
negative. The occurrence of local stability implies that the calculational 
procedure should be numerically convergent at the considered steps. Thus, even 
not knowing the exact solution of the problem and being unable to reach the 
limit of $k \ra \infty$, we can be sure that the local numerical convergence 
for finite $k$ is present.

\section{Free Energy}

In order to demonstrate that the self-similar approximation theory improves 
the results of optimized perturbation theory, it is instructive to consider 
the same problem of calculating the free energy (thermodynamic potential) of 
the model discussed in Sec. 7,
\be
\label{168}
 f(g) = - \ln\; Z(g) \qquad ( g > 0 ) \;  ,
\ee
with the statistical sum (\ref{57}). 

Following Sec. 7, we accept the initial Hamiltonian (\ref{59}) and define 
Hamiltonian (\ref{60}). Expanding the free energy (\ref{168}) in powers of 
the dummy parameter $\ep$, we have the sequence of approximants (\ref{62}). 
The control functions $\om_k(g)$ are defined by the optimization conditions 
(\ref{63}) and (\ref{64}), which give
\be
\label{169}
\om_k(g) = \left[\; 
\frac{1}{2} \; \left( 1 + \sqrt{1+12s_k g } \; \right) \; \right]^{1/2} \; ,
\ee
where
$$
s_1 =s_2 = 1 \; , \qquad s_3 =s_4 = 2.239674 \; .
$$ 

The rheonomic constraint (\ref{125}) takes the form
\be
\label{170}
F_0(g,\om_k(g)) = \ln \; \om_k(g) = f \; .
\ee
From here, we find the expansion function
\be
\label{171}
 g_k(f) = \frac{e^{2f}}{3s_k}\; \left( e^{2f} - 1 \right) \; .
\ee
The endomorphism (\ref{128}) reads as
\be
\label{172}
y_k(f) = f + \sum_{n=1}^k A_{kn} \al^n(f) \; ,
\ee
with the coefficients $A_{kn}$ given in Refs. 
\cite{Yukalov_25,Yukalov_62,Yukalov_65,Yukalov_66}, and where
\be
\label{173}
\al(f) = 1 - e^{-2f} \;  .
\ee
The cascade velocity (\ref{148}) becomes
\be
\label{174}
 v_k(f) = A_{k+1,k+1} \al^{k+1}(f) \; .
\ee
Taking the evolution integral (\ref{147}), with $t_k=1$, we come to the 
self-similar approximants
\be
\label{175}
  f^*_k(g) = F^*_k(g,\om_k(g) ) \; .
\ee

The accuracy of the approximations is described by the percentage errors
\be
\label{176}
 \ep^*_k \equiv \sup_g 
\left|\; \frac{f^*_k(g)-f(g)}{f(g)} \; \right|\times 100\% \; ,
\ee
where $f(g)$ is the exact numerical value of expression (\ref{168}). Here we 
have
$$
 \ep^*_1 = 3\%, \qquad \ep^*_2 = 2\%, \qquad \ep^*_3 = 0.1\% \;  .
$$

The map multipliers (\ref{155}) are
\be
\label{177}
\mu_k(f) = 
1 + 2 [\; 1 - \al(f) \;] \sum_{n=1}^k n A_{kn} \al^{n-1}(f) \; .
\ee
The coupling parameter $g$ pertains to the domain $[0,\infty)$, Then 
$f\in [0,\infty)$, and $\al(f)\in [0.1)$. The maximal map multiplier (\ref{157}) 
is found to satisfy the stability condition (\ref{158}).

\section{Fractal Transform}

As is explained in Sec. 4, control functions can be incorporated into a 
perturbative sequence either through initial conditions, or by means of the 
change of variables, or by a sequence transformation. In the above example 
of Sec. 14, we have considered the implantation of control functions into 
initial conditions. Now we shall study another way, when control functions 
are incorporated through a sequence transformation.  

Let us consider an asymptotic series
\be
\label{178}
f_k(x) = f_0(x) \left( 1 + \sum_{n=1}^k a_n x^n \right) \;  ,
\ee
in which $f_0(x)$ is a given function. Actually, it is sufficient to deal 
with the series
\be
\label{179}
f_k(x) =  1 + \sum_{n=1}^k a_n x^n  \;   .
\ee
To return to the case of series (\ref{178}), we just need to make the 
substitution
\be
\label{180}
 f_k(x) \longmapsto f_0(x) f_k(x) \;  .
\ee

Following the spirit of self-similarity, we can remember that the latter is 
usually connected with the power-law scaling and fractal structures 
\cite{Paladin_67,Kroger_68,Barnsley_69}. Therefore, it looks natural to 
introduce control functions through a fractal transform \cite{Barnsley_70}, 
say of the type \cite{Yukalov_24,Yukalov_25,Yukalov_31,Gluzman_32,Yukalov_33} 
\be
\label{181}
 F_k(x,s) = x^s f_k(x) \;  .
\ee
The inverse transformation is
\be
\label{182}
 f_k(x) = x^{-s} F_k(x,s) \;  .
\ee
With the series (\ref{179}), we have
\be
\label{183}
F_k(x,s) = x^s + \sum_{n=1}^k a_n x^{n+s} \;  .
\ee
As is mentioned in Sec. 4, the scaling relation (\ref{30}) is valid. The 
scaling exponent $s$ plays the role of a control parameter. 

In line with the self-similar approximation theory, we define the rheonomic 
constraint
\be
\label{184}
F_0(x,s) = x^s = f
\ee
yielding the expansion function
\be
\label{185}
 x(f) = f^{1/s} \;  .
\ee
The dynamic endomorphism becomes
\be
\label{186}
 y_k(f) = f + \sum_{n=1}^k a_n f^{1+n/s}\;  .
\ee
And the cascade velocity is
\be
\label{187}
 v_k(f) = y_k(f) - y_{k-1}(f) = a_k f^{1+n/s} \; .
\ee
What now remains is to consider the evolution integral.

\section{Self-Similar Root Approximants}

The differential equation (\ref{145}) can be rewritten in the integral form
\be
\label{188}
 \int_{y^*_{k-1}}^{y^*_k} \frac{df}{v_k(f)} = t_k \; .
\ee
Substituting here the cascade velocity (\ref{187}) gives the relation
\be
\label{189}
y^*_k(f) = 
\left\{ \left[ \; y^*_{k-1}(f)\; \right]^{1/m_k} + A_k \right\}^{m_k} \; ,
\ee
where
$$
m_k \equiv -\; \frac{s_k}{k} \; , \qquad 
A_k \equiv \frac{a_kt_k}{m_k}\;   .
$$

Accomplishing the inverse transformation (\ref{182}) leads to the equation
\be
\label{190}
f^*_k(x) = x^{-s} y^*_k(f) \qquad ( f = x^s ) \; .
\ee
The explicit form of the latter is the recurrent relation
\be
\label{191}
f^*_k(x) = 
\left\{ \left[ \; f^*_{k-1}(x)\; \right]^{1/m_k} + A_k x^k\right\}^{m_k} \;   .
\ee
Using the notation
\be
\label{192}
 n_j \equiv \frac{m_j}{m_{j+1}} = \frac{(j+1)s_j}{js_{j+1}} \qquad
( j = 1,2,\ldots, k-1 )  
\ee
and iterating this relation $k-1$ times results in the self-similar root 
approximant
\be
\label{193}
f^*_k(x) = \left( \left( (1 + A_1 x)^{n_1} + A_2 x^2 \right)^{n_2} + 
\ldots + A_k x^k \right)^{m_k} \; .
\ee
This approximant is convenient for the problem of interpolation, where one can 
meet different situations. 

\vskip 2mm
(i) The $k$ coefficients $a_n$ of the asymptotic expansion (\ref{179}) up 
to the $k$-th order are known and the exponent $\beta$ of the large-variable 
behavior of the sought function is available, where
\be
\label{194}
f(x) \simeq B x^\bt \qquad ( x \ra \infty) \; ,
\ee
although the amplitude $B$ is not known. Then, setting the control functions 
$s_j=s$, from Eq. (\ref{192}), we have
\be
\label{195}
 n_j = \frac{j+1}{j} \qquad ( j = 1,2, \ldots, k-1 ) \;  ,
\ee
and the root approximant (\ref{193}) becomes
\be
\label{196}
f^*_k(x) = \left( \left( \left( ( 1 + A_1 x)^2 + A_2 x^2 \right)^{3/2} + 
A_3 x^3 \right)^{4/3} + \ldots  + A_k x^k \right)^{m_k} \; .
\ee
For large variables $x$, the latter behaves as
\be
\label{197}
 f^*_k(x) \simeq B_k x^{\bt_k} \qquad ( x \ra \infty) \;  ,
\ee
with the amplitude 
\be
\label{198}
 B_k =  \left( \left( \left( A_1^2 + A_2 \right)^{3/2} + A_3 \right)^{4/3} +
\ldots + A_k \right)^{m_k}  
\ee
and exponent 
\be
\label{199}
  \bt_k = k m_k \; .
\ee

Equating $\beta_k$ to the known exponent $\beta$, we find the root exponent
\be
\label{200}
 m_k = \frac{\bt}{k} \qquad ( \bt_k = \bt) \;  .
\ee
All parameters $A_n$ can be found from the comparison of the initial 
series (\ref{179}) with the small-variable expansion of the root approximant 
(\ref{196}), 
\be
\label{201}
  f^*_k(x) \simeq  f_k(x) \qquad ( x \ra 0) \; ,
\ee
which is called the accuracy-trough-order procedure. Knowing all $A_n$, we 
obtain the large-variable amplitude $B_k$. 

\vskip 2mm
(ii) The $k$ coefficients $a_n$ of the asymptotic expansion (\ref{179}) up 
to the $k$-th order are available and the amplitude $B$ of the large-variable 
behavior of the sought function is known, but the large-variable exponent 
$\bt$ is not known. Then the parameters $A_n$ again are defined through the 
accuracy-through-order procedure (\ref{201}). Equating the amplitudes $B_k$ 
and $B$ results in the exponent
\be
\label{202}
 m_k = \frac{\ln B}{\ln(((A_1^2+A_2)^{3/2}+A_3)^{4/3}+\ldots+A_k)} \qquad (B_k = B) \; .
\ee

\vskip 2mm
(iii) The $k$ coefficients $a_n$ of the asymptotic expansion (\ref{179}) are 
known and the large-variable behavior (\ref{194}) is available, with both the 
amplitude $B$ and exponent $\bt$ known. Then, as earlier, the parameters $A_n$ 
are defined from the accuracy-through-order procedure and the exponent $m_k$ 
is given by Eq. (\ref{200}). The amplitude $B_k$ can be found in two ways, from 
expression (\ref{198}) and equating $B_k $ and $B$. The difference between the 
resulting values defines the accuracy of the approximant.    

\vskip 2mm
(iv) The $k$ terms of the large-variable behavior are given,
\be
\label{203}
 f(x) \simeq \sum_{n=1}^k b_n x^{\bt_n} \qquad ( x \ra \infty) \;  ,
\ee
where $b_1\neq 0$, $\bt_1\neq 0$, and the powers $\bt_n$ are arranged in the 
descending order,
\be
\label{204}
 \bt_n > \bt_{n+1} \qquad ( n = 1,2,\ldots, k-1 ) \;  .
\ee
Then considering the root approximant (\ref{193}) for large $x\ra\infty$, 
and comparing this expansion with the asymptotic form (\ref{203}) we find 
all parameters $A_n$ expressed through the coefficients $b_n$, and the 
large-variable internal exponents are 
\be
\label{205}
n_j = \frac{j+1}{j} + \frac{1}{j} \; (\bt_{k-j+1} - \bt_{k-j} ) \qquad
( j = 1,2,\ldots, k-1) \;  ,
\ee
while the external exponent is
\be
\label{206}
m_k =\frac{\bt_1}{k} \;  .
\ee

\vskip 2mm
It is important to mention that the external exponent $m_k$ can be defined 
even without knowing the large-variable behavior of the sought function. This 
can be done by treating $m_k$ as a control function defined by an optimization 
condition from Sec. 6. This method has been suggested in Ref. \cite{Yukalov_31}. 

Notice that when it is more convenient to deal with the series for large 
variables, it is always possible to use the same methods as described above by 
transferring the large-variable expansions into small-variable ones by means 
of the change of the variable $z=1/x$. 

Numerous applications of the self-similar root approximants to different 
problems are discussed in Refs. 
\cite{Yukalov_24,Yukalov_25,Yukalov_71,Gluzman_72,Yukalov_73,Yukalov_74,Yukalov_75}.

\section{Self-Similar Nested Approximants}

It is possible to notice that the series
\be
\label{207}
f_k(x) = 1 + \sum_{n=1}^k a_n x^n
\ee
can be represented as the sequence
$$
f_k(x) = 1 + \vp_1(x) \; , \qquad \vp_1(x) = a_1 x\;( 1 + \vp_2(x) ) \; 
$$
\be
\label{208}
\vp_2(x) = \frac{a_2}{a_1} \; x \; (1 + \vp_3(x) ) \; , \qquad 
\vp_3(x) = \frac{a_3}{a_2} \;  x \; ( 1 + \vp_4(x) ) \;  ,
\ee
etc., through 
\be
\label{209}
 \vp_j(x) = \frac{a_j}{a_{j-1}} \; x \; (1 + \vp_{j+1}(x) ) \qquad
(j = 1,2 \ldots,k-1 ) \;   ,
\ee
up to the last term
\be
\label{210}
 \vp_k(x) = \frac{a_k}{a_{k-1}} \; x \; .
\ee

Applying the self-similar renormalization at each order of the sequence, 
considering $\varphi_j$ as variables, we obtain the renormalized sequence
\be
\label{211}
 f^*_k(x) = \left( 1 + b_1 \vp^*_1(x) \right)^{n_1} \;  , 
\qquad
\vp^*_j(x) = \frac{a_j}{a_{j-1}} \; x \; 
\left( 1 + b_{j+1}\vp^*_{j+1}(x) \right)^{n_{j+1}} \; ,
\ee
in which
$$
b_j =\frac{t_j}{n_j} \; , \qquad n_j = - s_j   \qquad 
(j = 1,2 \ldots,k-1 ) \;   .
$$
Using the notation
\be
\label{212}
 A_j =\frac{a_j}{a_{j-1}} \; b_j = \frac{a_jt_j}{a_{j-1}n_j} \; ,
\ee
we come to the self-similar nested approximant
\be
\label{213}
 f^*_k(x) = \left( 1 + A_1x\left( 1 + A_2x \ldots ( 1 + A_k x)^{n_k}
\right)^{n_{k-1}} \ldots \right)^{n_1} \; .
\ee

For large $x$, this gives
\be
\label{214}
f^*_k(x) \simeq B_k x^\bt_k \qquad ( x \ra \infty) \;  ,
\ee
with the amplitude
\be
\label{215}
B_k = A_1^{n_1} A_2^{n_1n_2} A_3^{n_1n_2n_3} \ldots 
A_k^{n_1n_2n_3\ldots n_k}
\ee
and the exponent
\be
\label{216}
 \bt_k = n_1 +n_1n_2 + n_1n_2n_3 + \ldots + n_1n_2n_3\ldots n_k \;  .
\ee

If we change the notation for the external exponent to
\be
\label{217}
 m_k \equiv n_1 \;  ,
\ee
and keep the internal exponents constant,
\be
\label{218}
 n_j = m \qquad ( j = 2,3,\ldots, k) \;  ,
\ee
then the large-variable exponent becomes
\be
\label{219}
 \bt_k = \frac{1-m^k}{1-m} \; m_k \;  .
\ee

When the exponent $\beta$ of the large-variable behavior is known, where
\be
\label{220}
 f(x) ~ \propto ~ x^\bt \qquad ( x \ra \infty) \;  ,
\ee
then, setting $\beta_k = \beta$, gives
\be
\label{221}
 m_k = \frac{1-m}{1-m^k}\; \bt   \qquad ( \bt_k = \bt)  \;  .
\ee
   
The parameter $m$ should be defined so that to provide numerical convergence 
for the sequence $\{f^*_k(x)\}$. For instance, if $m = 1$, then using the 
asymptotic form
$$
m^k \simeq 1 - (1-m) k \qquad ( m \ra 1) \;   ,   
$$
we get 
\be
\label{222}
 m_k = \frac{\bt}{k} \qquad ( m = 1) \;  .
\ee
In the latter case, the nested approximant (\ref{213}), with the notation
$$
 D_n \equiv \prod_{j=1}^n A_j \;  ,
$$
becomes
$$
f^*_k(x) = 
\left( 1 + D_1 x + D_2 x^2 + D_3 x^3 + \ldots + D_k x^k \right)^{m_k} \; .   
$$
The same form can be obtained by setting in the root approximant (\ref{193}) 
all internal exponents $n_j = 1$.

The external exponent $m_k$ can also be defined by resorting to the optimization 
conditions of Sec. 6. Several applications of the nested approximants are given 
in \cite{Gluzman_76}.

\section{Self-Similar Exponential Approximants}

When it is expected that the behavior of the sought function is rather 
exponential, but not of power law, then in the nested approximants of the 
previous section, we can sent $n_j\ra\infty$, hence $b_j\ra 0$ and $A_j\ra 0$. 
This results in the self-similar exponential approximants \cite{Yukalov_77}
\be
\label{223}
 f^*_k(x) = \exp\left( C_1 x \exp\left( C_2 x \exp\left( C_3 x \ldots 
\exp(C_k x) \right) \right) \right) \; ,
\ee
in which
\be
\label{224}
 C_n = \frac{a_n}{a_{n-1}} \; t_n \qquad ( n = 1,2, \ldots, k ) \;  .
\ee
The parameters $t_n$ are to be defined from additional conditions 
\cite{Yukalov_24,Yukalov_25}, so that the sequence of the approximants be 
convergent. It is often sufficient to set $t_n=1/n$. This expression appears 
as follows. By its meaning, $t_n$ is the effective time required for reaching 
a fixed point from the previous step. Accomplishing $n$ steps takes time of 
order $n t_n$. The minimal time corresponds to one step. Equating $n t_n$ and 
one gives $t_n = 1/n$. Some other ways of defining the control parameters $t_n$ 
are considered in Refs. \cite{Yukalov_24,Yukalov_25,Yukalov_77}.

\section{Self-Similar Factor Approximants}

By the fundamental theorem of algebra \cite{Lang_78}, a polynomial of any degree of one 
real variable over the field of real numbers can be split in a unique way into a product
of irreducible first-degree polynomials over the field of complex numbers. This means 
that series (\ref{207}) can be represented in the form
\be
\label{225}
  f_k(x) = \prod_{j=1}^k ( 1 +b_j x) \;  ,
\ee
with the coefficients $b_j$ expressed through $a_n$. Applying the self-similar 
renormalization procedure to each of the factors in turn results in the 
self-similar factor approximants \cite{Yukalov_79,Gluzman_80,Yukalov_81}
\be
\label{226}
 f^*_k(x) = \prod_{j=1}^{N_k} ( 1 + A_j x)^{n_j} \; ,
\ee
where
\begin{eqnarray}
\nonumber
N_k = \left\{ \begin{array}{ll}
k/2 \; , ~ & ~ k = 2,4,6,\ldots \\
(k+1)/2 \; , ~ & ~ k = 1,3,5,\ldots
\end{array} \right. .
\end{eqnarray}

The control parameters $A_j$ and $n_j$ are defined by the accuracy-through-order
procedure by equating the like order terms in the expansions $f^*_k(x)$ and 
$f_k(x)$,
\be
\label{227}
 f^*_k(x) \simeq f_k(x) \qquad ( x \ra 0 ) \;  .
\ee
In the present case, it is more convenient to compare the corresponding 
logarithms
\be
\label{228}
\ln\; f^*_k(x) \simeq \ln \; f_k(x) \qquad ( x \ra 0 ) \;  .
\ee
This leads to the system of equations
\be
\label{229}
 \sum_{j=1}^{N_k} n_j A_j^n = D_n  \qquad (n = 1,2,\ldots,k) \;  ,
\ee
in which
\be
\label{230}
D_n \equiv \frac{(-1)^{n-1}}{(n-1)!} \; \lim_{x\ra 0} \;
\frac{d^n}{dx^n}\; \ln \; \left( 1 + \sum_{m=1}^n a_m x^m \right) \; .
\ee
This system of equations enjoys a unique (up to enumeration permutation) 
solution for all $A_j$ and $n_j$ when $k$ is even, and when $k$ is odd, one 
of $A_j$ can be set to one \cite{Yukalov_25,Yukalov_82}.

At large values of the variable, we have
\be
\label{231}
 f^*_k(x) \simeq B_k x^{\bt_k}  \qquad (x \ra \infty ) \;  ,
\ee
where the amplitude and the large-variable exponent are
\be
\label{232}
B_k = \prod_{j=1}^{N_k} A_j^{n_j} \; , \qquad 
\bt_k = \sum_{j=1}^{N_k} n_j \;  .
\ee

If the large-variable exponent is known, for instance from scaling arguments, 
so that
\be
\label{233}
 f(x) ~ \propto ~ x^\bt \qquad ( x \ra \infty) \; ,
\ee
then equating $\beta_k$ and $\beta$ imposes on the exponents of the factor 
approximant the constraint
\be
\label{234}
 \bt_k = \sum_{j=1}^{N_k} n_j = \bt \;  .
\ee

The self-similar factor approximants have been used for a variety of problems, 
as can be inferred from 
Refs. \cite{Yukalov_25,Yukalov_79,Gluzman_80,Yukalov_81,Yukalov_82,Yukalov_99}.

\section{Self-Similar Combined Approximants}

It is possible to combine different types of self-similar approximants as well 
as these approximants and other kinds of approximations.

\subsection{Different Types of Approximants}

Suppose we are given a small-variable asymptotic expansion
\be
\label{235}
 f_k(x) = \sum_{j=0}^{k} a_j x^j \qquad  ( x \ra 0 )  \; ,
\ee
which we plan to convert into a self-similar approximation. At the same time, 
we suspect that the behavior of the sought function is quite different at small 
and at large variables. In such a case, we can combine different types of 
self-similar approximants in the following way. We take in series (\ref{235}) 
several initial terms,
\be
\label{236}
 f_n(x) = \sum_{j=0}^{n} a_j x^j \qquad ( n < k )   
\ee
and construct of them a self-similar approximant $f^*_n(x)$. Then we define 
the ratio
\be
\label{237}
C_{k/n}(x) \equiv \frac{f_k(x)}{f^*_n(x)}
\ee
and expand the latter in powers of $x$ as
\be
\label{238}
C_{k/n}(x) = 1 + \sum_{j=n+1}^{k} b_j x^j \qquad ( x \ra 0 )  \;  .
\ee
Constructing a self-similar approximant $C^*_{k/n}(x)$, we obtain the combined 
approximant
\be
\label{239}
  f^*_k(x) = f^*_n(x) C^*_{k/n}(x) \; .
\ee

The approximants $f^*_n(x)$ and $C^*_{k/n}(x)$ can be represented by 
different forms of self-similar approximants. For example, it is possible 
to define $f^*_n(x)$ as a root approximant, while $C^*_{k/n}(x)$ as a factor 
or exponential approximant, depending on the expected behavior of the sought 
function \cite{Gluzman_83}.

\subsection{Self-Similar Pad\'{e} Approximants}

Instead of two different self-similar approximants, it is possible, after 
constructing a self-similar approximant $f^*_n(x)$, to transform the remaining 
part (\ref{238}) into a Pad\'{e} approximant $P_{M/N}(x)$, with $M+N=k-n$, so 
that 
\be
\label{240}
 P_{M/N}(x) \simeq C_{k/n}(x) \qquad ( x \ra 0 )  \;   .
\ee
The result is the self-similarly corrected Pad\'{e} approximant, or briefly, 
the self-similar Pad\'{e} approximant \cite{Gluzman_84,Gluzman_85,Gluzman_86}
\be
\label{241}
 f^*_k(x) = f^*_n(x) \; P_{M/N}(x) \;  .
\ee

The advantage of this type of approximants is that they can correctly take into 
account irrational behavior of the sought function, described by the self-similar 
approximant $f^*_n(x)$, as well as the rational behavior represented by the 
Pad\'{e} approximant $P_{M/N}(x)$. 

Note that Pad\'{e} approximants (\ref{8}) actually are a particular case of 
the factor approximants (\ref{226}), where $M$ factors correspond to $n_j = 1$ 
and $N$ factors, to $n_j = -1$. This is because the Pad\'{e} approximants can be 
represented as
$$
 P_{M/N}(x) = 
a_0 \prod_{m=1}^M ( 1 + A_m x) \; \prod_{n=1}^N (1+C_n x)^{-1} \; .
$$

\subsection{Self-Similar Borel Summation}

It is possible to combine self-similar approximants with the method of Borel 
summation. According to this method, for a series (\ref{235}), one can define 
\cite{Hardy_9,Weinberg_87} the Borel-Leroy transform
\be
\label{242}
 B_k(t,u) \equiv \sum_{n=0}^k \frac{a_n}{\Gm(n+1+u)} \; t^n \;  ,
\ee
where $u$ is chosen so that to improve convergence. The series (\ref{242}) 
can be summed using one of the self-similar approximations, and converting 
$u$ into a control parameter $u_k$, thus getting $B^*_k(t,u_k)$. Then the 
self-similar Borel-Leroy summation yields the approximant
\be
\label{243}
 f^*_k(x) = \int_0^\infty e^{-t} t^{u_k} B^*_k(tx,u_k) \; dt \;  .
\ee
The case of the standard Borel summation corresponds to $u_k = 0$. Then the 
self-similar Borel summation gives
\be
\label{244}
f^*_k(x) = \int_0^\infty e^{-t}  B^*_k(tx) \; dt \;    .
\ee

\vskip 2mm

In addition to the considered above combinations of different summation methods, one 
can use other combinations. For example, the combination of exponential approximants and
continued fractions has been employed \cite{Abhignan_118}.

\section{Self-Similar Data Extrapolation}

One often meets the following problem. There exists an ordered dataset
\be
\label{245}
 \{ f_n: ~ n = 1,2,\ldots,k\} 
\ee
labeled by the index $n$, and one is interested in the possibility 
of predicting the values $f_{k+p}$ outside this dataset. The theory of 
self-similar approximants suggests a solution to this problem 
\cite{Yukalov_25,Yukalov_88}.

Let us consider several last datapoints, for instance the last three points
\be
\label{246}
\{ g_0 \equiv f_{k-2}\; , ~ g_1 \equiv f_{k-1}\; , ~ g_2 \equiv f_k \} \;  .
\ee
How many datapoints one needs to take depends on the particular problem 
considered. For the explicit illustration of the idea, we take three 
datapoints. The chosen points can be connected by a polynomial spline, in 
the present case, by a quadratic spline   
\be
\label{247}
g(t) = a + bt + ct^2 
\ee
defined so that 
\be
\label{248}
  g(0) = g_0 = f_{k-2} \; , \qquad g(1) = g_1 = f_{k-1} \; , \qquad
g(2) = g_2 = f_k \; .
\ee
From this definition it follows
$$
a = f_{k-2} \; , \qquad 
b = - \; \frac{1}{2} \; ( f_k - 4f_{k-1}+3f_{k-2} ) \; , \qquad
c = \frac{1}{2} \; ( f_k - 2f_{k-1}+f_{k-2} ) \; .
$$

Treating polynomial (\ref{247}) as an expansion in powers of $t$ makes it 
straightforward to employ self-similar renormalization, thus, obtaining a 
self-similar approximant $g^*(t)$. For example, resorting to factor approximants, 
we get
\be
\label{249}
 g^*(t) = a ( 1 + At)^m \;  ,
\ee
with the parameters
$$
 A = \frac{b^2-ac}{ab} \; , \qquad m = \frac{b^2}{b^2-ac} \;  .
$$
The approximants $g^*(t)$, with $t \geq 2$ provide the extrapolation of the 
initial dataset. The nearest to the dataset extrapolation point can be estimated 
as
\be
\label{250}
 g^* = \frac{1}{2}\; \left[ \; g^*(2) + g^*(3) \; \right] \; .
\ee 
  
This method can also be used for improving the convergence of the sequence of 
self-similar approximants. Then the role of datapoints $f_k$ is played by the 
self-similar approximants $f^*_k(x)$. In that case, all parameters $a = a(x)$, 
$b = b(x)$, $c = c(x)$, as well as $A = A(x)$ and $m = m(x)$ become control 
functions. This method of data extrapolation has been used for several problems, 
such as predictions for time series and convergence acceleration 
\cite{Yukalov_25,Yukalov_88,Yukalov_89}.

\section{Self-Similar Diff-Log Approximants}

There is a well known method employed in statistical physics called diff-log 
transformation \cite{He_90,Gluzman_91}. This transformation for a function 
$f(x)$ is
\be
\label{251}
 D(x) \equiv \frac{d}{dx} \; \ln \;f(x) \;  .
\ee
The inverse transformation, assuming that the function $f(x)$ is normalized 
so that 
\be
\label{252}
 f(0) = 1 \; ,
\ee
reads as
\be
\label{253}
 f(x) = \exp\left\{ \int_0^x D(t) \; dt \right\} \;  .
\ee

When we start with an asymptotic expansion
\be
\label{254}
 f_k(x) = 1 + \sum_{n=1}^k a_n x^n \;  ,
\ee
the diff-log transformation gives
\be
\label{255}
 D_k(x) = \frac{d}{dx} \; \ln \; f_k(x) \;  .
\ee
Expanding the latter in powers of $x$ yields
\be
\label{256}
 D_k(x) \simeq \sum_{n=0}^k b_n x^n \qquad ( x \ra 0 ) \;  ,
\ee
with the coefficients $b_n$ expressed through $a_n$. This expansion can be 
summed by one of the self-similar methods giving $D_k^*(x)$. Involving the 
inverse transformation (\ref{253}) results in the self-similar diff-log 
approximants
\be
\label{257}
f^*_k(x) = \exp\left\{ \int_0^x D^*_k(t) \; dt \right\} \;   .
\ee
A number of applications of the diff-log transformation can be found in Refs. 
\cite{Yukalov_74,Gluzman_86,Gluzman_91}, where it is shown that the 
combination of the diff-log transform with self-similar approximants gives 
essentially more accurate results than the diff-log Pad\'e method.

\section{Critical Behavior}

One says that a function $f(x)$ experiences critical behavior at a critical 
point $x_c$, when this function at that point either tends to zero or to 
infinity. It is possible to distinguish two cases, when the critical behavior 
occurs at infinity, and when at a finite critical point. These two cases will 
be considered below separately.

\subsection{Critical Point at Infinity}

If the critical behavior happens at infinity, the considered function behaves 
as  
\be
\label{258}
 f(x) \simeq B x^\bt \qquad ( x \ra \infty ) \;  .
\ee
Then the diff-log transform tends to the form
\be
\label{259}
D(x) \simeq \frac{\bt}{x} \qquad  ( x \ra \infty ) \;  .
\ee
Here $B$ is a critical amplitude, while $\beta$ is a critical exponent. 

The critical exponents have a special interest for critical phenomena. If we 
are able to define a self-similar approximation $f_k^*(x)$ directly to the 
studied function $f(x)$, then the critical exponent can be found from the limit
\be
\label{260}
\bt_k = \lim_{x\ra\infty} \; \frac{\ln f^*_k(x)}{\ln x} \;   .
\ee
Otherwise, it can be obtained from the equivalent form
\be
\label{261}
 \bt_k = \lim_{x\ra\infty} \; x D^*_k(x) \; ,
\ee
where a self-similar approximation for the diff-log transform $D_k^*(x)$ is 
needed.

The convenience of using the representation (\ref{261}) is in the possibility 
of employing a larger arsenal of different self-similar approximants. Of 
course, the factor approximants can be involved in both the cases. However, 
the root and nested approximants require the knowledge of the large-variable 
exponent of the sought function, which is not always available. On the 
contrary, the large-variable behavior of the diff-log transform (\ref{259}) 
is known. Therefore for constructing a self-similar approximation for the 
diff-log transform, we can resort to any type of self-similar approximants.

It is necessary to mention that the root and nested approximants can be defined, 
without knowing the large-variable behavior, by invoking optimization conditions 
of Sec. 6 prescribing the value of the external exponent $m_k$, as is explained 
in Ref. \cite{Yukalov_31}. However this method becomes rather cumbersome for 
high-order approximants.

\subsection{Finite Critical Point}

If the critical point is located at a finite $x_c$ that is in the interval 
$(0, \infty)$, then
\be
\label{262}
 f(x) \simeq B ( x_c - x)^\bt \qquad ( x \ra x_c -0 ) \;  .
\ee
Here the diff-log transform behaves as
\be
\label{263}
D(x) \simeq -\; \frac{\bt}{ x_c - x} \qquad ( x \ra x_c -0 ) \;    .
\ee

Again, the critical exponent can be derived from the limit
\be
\label{264}
 \bt_k = \lim_{x\ra x_c-0} \; \frac{\ln f_k^*(x)}{\ln(x_c-x)} \;  ,
\ee
provided a self-similar approximant $f_k^*(x)$ is constructed. But it may 
happen that the other form
\be
\label{265}
\bt_k = \lim_{x\ra x_c-0} \; (x-x_c) D_k^*(x)
\ee
is more convenient, where a self-similar approximant for the diff-log 
transform $D_k^*(x)$ is easier to find. This is because the nearest to zero 
pole of $D_k^*(x)$ defines a critical point $x_c$, while the residue (\ref{265}) 
yields a critical exponent. 

Note that by the change of the variable the problem of a finite critical point 
can be reduced to the case of critical behavior at infinity. For instance one 
can use the change of the variable $z = x/(x_c -x)$ or any other change of the 
variable mapping the interval $[0,x_c)$ to $[0,\infty)$. Numerous examples of 
applying the diff-log transform, accompanied by the use of self-similar 
approximants, are presented in Refs. \cite{Yukalov_74,Gluzman_86,Gluzman_91}, 
where it is also shown that this method essentially outperforms the diff-log 
Pad\'{e} variant.

\section{Non-Power-Law Behavior}

In the previous sections we were mainly keeping in mind a kind of power-law 
behavior of considered functions at large variables. Now it is useful to make 
some comments on the use of the described approximation methods for other types 
of behavior. The most often met types of behavior that can occur at large 
variables are the exponential and logarithmic behavior. Below we show that 
the developed methods of self-similar approximants can be straightforwardly 
applied to any type of behavior.

\subsection{Exponential Behavior}   

The exponential behavior with respect to time happens in many mathematical 
models employed for describing the growth of population, mass of biosystems, 
economic expansion, financial markets, various relaxation phenomena, etc.
\cite{Zeide_92,Day_93,Yukalov_94,Yukalov_95,Yukalov_96,Gluzman_97}.
 
When a sought function at a large variable displays exponential behavior, there 
are several ways of treating this case. First of all, this kind of behavior can 
be treated by self-similar exponential approximants of Sec. 18. The other way 
is to resort to diff-log approximants of Sec. 22 or, simply, to consider the 
logarithmic transform
\be
\label{266}
 L(x) \equiv \ln f(x) \;  .
\ee
If the sought function at large variable behaves as
\be
\label{267}
 f(x) \simeq B\exp(\gm x) \qquad ( x \ra \infty) \;  ,
\ee
then
\be
\label{268}
L(x) \simeq \gm x \; ,  \qquad D(x) \simeq \gm \qquad ( x \ra \infty) \; .
\ee
Therefore the function behaves as
\be
\label{269}
 f(x) \simeq B\exp\{L( x) \} \qquad ( x \ra \infty) \;  .
\ee
 
Keeping in mind the asymptotic series (\ref{254}), we have
\be
\label{270}
 L_k(x) = \ln f_k(x) \;  ,
\ee
which can be expanded in powers of $x$ giving
\be
\label{271}
L_k(x) \simeq \sum_{n=0}^k c_n x^n \qquad ( x \ra 0 ) \;   .
\ee
This is to be converted into a self-similar approximant $L_k^*(x)$, after 
which we get the answer
\be
\label{272}
 f_k^*(x) = \exp\{ L_k^*(x) \} \;  .
\ee
  
Moreover, the small-variable expansion of an exponential function can be 
directly and exactly represented through self-similar factor approximants 
\cite{Yukalov_82}. Really, let us consider the exponential function
\be
\label{273}
 f(x) = e^x \;  .
\ee
Assume that we know solely the small-variable asymptotic expansion
\be
\label{274}
f_k(x) = \sum_{n=0}^k \frac{x^n}{n!} \qquad ( x \ra 0 ) \; ,
\ee
which is used for constructing factor approximants. In the lowest, second, 
order we have
$$
f_2^*(x) = \lim_{A\ra 0} ( 1 + Ax)^{1/A} = e^x \;  .
$$
In the third order, we find
$$
f_3^*(x) = 
\lim_{A\ra 0} ( 1 + x)^{A/(1-A)} ( 1 + Ax)^{1/A(1-A)} = e^x \; ,
$$
and, similarly, in all other orders. So that the self-similar factor 
approximants of all  orders reproduce the exponential function exactly:
\be
\label{275}
 f^*_k(x) = e^x \qquad ( k \geq 2) \;  .
\ee
Some other more complicated functions, containing exponentials, also can be 
well approximated by factor approximants \cite{Yukalov_98}.

\subsection{Logarithmic Behavior}

When there is suspicion that the sought function exhibits logarithmic behavior 
at large variables, it is reasonable to act by analogy with the previous 
subsection, but now defining the exponential transform
\be
\label{276}
 E(x) \equiv \exp\{ f(x) \} \;  .
\ee
For the asymptotic series (\ref{254}), we have
\be
\label{277}
E_k(x) \equiv \exp\{ f_k(x) \} \;   ,
\ee
whose expansion in powers of $x$ produces
\be
\label{278}
 E_k(x) = \sum_{n=0}^k b_n x^n \qquad ( x \ra 0 ) \;  .
\ee
This can be converted into a self-similar approximation $E_k^*(x)$, so that 
the final answer becomes
\be
\label{279}
 f_k^*(x) = \ln E_k^*(x) \;  .
\ee

As an example, let us consider the function
\be
\label{280}
f(x) = 1 + \ln\left( \frac{1+\sqrt{1+x}}{2} \right) \; ,
\ee
with the logarithmic behavior at large variables,
\be
\label{281}
 f(x) \simeq 0.5\ln x \qquad ( x \ra \infty ) \; .
\ee
This function has the expansion
\be
\label{282}
 f_k(x) =\sum_{n=0}^k a_n x^n \qquad ( x \ra 0 ) \;  ,
\ee
with the coefficients
$$
a_0 = 1 \; , \qquad a_1 = \frac{1}{4} \; , \qquad 
a_2 = - \; \frac{3}{32} \; , \qquad a_3 = \frac{5}{96} \; , 
$$
$$
a_4 = -\;\frac{35}{1024} \; , \qquad a_5 = \frac{63}{2560} \; , 
\qquad a_6 = - \; \frac{77}{4096} \; , \qquad \ldots \; .
$$
Its exponential transform leads to the series (\ref{278}), with the 
coefficients
$$
b_0 = e \; , \qquad b_1 = \frac{1}{4}\; e \; , \qquad 
b_2 = - \; \frac{1}{16}\; e \; ,  \qquad b_3 = \frac{1}{32}\; e \; ,
$$
$$
b_4 = -\;\frac{5}{256}\; e \; , \qquad b_5 = \frac{7}{512}\; e \; , 
\qquad b_6 = - \; \frac{21}{2048}\; e \; , \qquad \ldots \;.
$$
Defining factor approximants $E_k^*(x)$, we obtain the approximants (\ref{279}), 
whose large-variable behavior is of correct logarithmic form
\be
\label{283}
 f_k^*(x) \simeq B_k \ln x \qquad ( x \ra \infty ) \;  ,
\ee
with the amplitudes $B_k$,
$$
B_2 = 0.333 \; , \qquad B_4 = 0.4 \; , \qquad B_6 = 0.429 \; ,
$$
$$
 B_8 = 0.444 \; , \qquad B_{10} = 0.456 \; , \qquad B_{12} = 0.462 \; ,
\qquad \ldots \; ,
$$ 
converging to the exact value $0.5$.

\section{Critical Temperature Shift}

Here we show how the described methods can be used for calculating the critical 
temperature relative shift caused by interactions in an $N$-component scalar 
field theory in three dimensions. The interactions can be characterized by the 
gas parameter
\be
\label{284}
 \gm \equiv \rho^{1/3} a_s \;  ,
\ee
in which $\rho$ is particle density and $a_s$, s-wave scattering length. This 
shift is defined as
\be
\label{285}
 \frac{\Dlt T_c}{T_0} \equiv \frac{T_c-T_0}{T_0} \;  ,
\ee
where $T_0$ is the critical temperature in the free field with $\gm=0$, 
while $T_c$ is the critical temperature for nonzero $\gm$. For example, the 
critical temperature of the $2$-component free field 
\be
\label{286}
T_0 = \frac{2\pi}{m} \; 
\left[ \; \frac{\rho}{\zeta(3/2)} \; \right]^{2/3}
\ee
is the point of the Bose-Einstein condensation of ideal gas. Here $m$ is the 
mass of a boson, and the Boltzmann and Planck constants are set to one. For weak 
interactions, the temperature shift has been shown \cite{Baym_100,Baym_101} to 
have the form
\be
\label{287}
\frac{\Dlt T_c}{T_0} \simeq c_1 \gm \qquad ( \gm \ra 0 ) \;  ,
\ee
where the coefficient $c_1$ needs to be calculated.  

This coefficient can be found in the loop expansion 
\cite{Kastening_102,Kastening_103,Kastening_104} producing asymptotic series in 
powers of the variable
\be
\label{288}
 x = (N + 2) \;\frac{ \lbd_{eff} }{ \sqrt{\mu_{eff} } } \;  ,
\ee
where $N$ is the number of components, $\lambda_{eff}$, effective coupling, and 
$\mu_{eff}$, effective chemical potential. The series in seven loops reads as
\be
\label{289}
 c_1(x) \simeq \sum_{n=1}^5 a_n x^n \qquad ( x \ra 0 ) \; ,
\ee
whose coefficients for several $N$ are listed in Table 1.

However, at the critical point, the effective chemical potential tends to 
zero, hence the variable $x$ tends to infinity. Thus we come to the necessity 
of finding the series (\ref{289}) for $x \ra \infty$. The direct application 
of the limit $x \ra \infty$ to this series of course has no sense. We use the 
self-similar factor approximants of Sec. 19, defining the approximants $f^*_k(x)$ 
for $c_1(x)$, with keeping in mind that $c_1$ is finite, so that $\bt_k=0$. Then 
the approximants for the sought limit are
\be
\label{290}
 f^*_k(\infty) = a_1 \prod_{i=1}^{N_k} A_i^{n_i} ~ \longmapsto ~ c_1 \; .
\ee
The convergence is accelerated by quadratic splines, as is explained in Sec. 21 
and in Refs. \cite{Yukalov_89,Yukalov_105}. The results are displayed in Table 2, 
where they are compared with Monte Carlo simulations 
\cite{Kashurnikov_106,Arnold_107,Arnold_108,Sun_109}. The agreement of the 
latter with the values calculated by means of the self-similar approximants 
is very good.

\section{Critical Exponents}

Calculation of critical exponents is one of the most important problems in the 
theory of phase transitions. Here we show how the critical exponents can be 
calculated by using self-similar factor approximants applied to the asymptotic 
series in powers of the $\ep=4-d$, where $d$ is space dimensionality. We shall 
consider the $O(N)$  $\varphi^4$ field theory in $d = 3$. The definition of the 
critical exponents can be found in reviews \cite{Yukalov_25,Pelissetto_110}.

One usually derives the so-called epsilon expansions for the exponents $\eta$, 
$\nu^{-1}$, and $\omega$. The other exponents can be obtained from the scaling 
relations
\be
\label{291}
\al= 2-\nu d \; , \qquad \bt = \frac{\nu}{2}\; ( d - 2 +\eta) \; , 
\qquad
\gm = \nu (2-\eta) \; , \qquad \dlt = \frac{d+2-\eta}{d-2+\eta} \;  .
\ee
In three dimensions, one has
\be
\label{292}
\al= 2- 3\nu \; , \qquad \bt = \frac{\nu}{2}\; ( 1 +\eta) \; , 
\qquad
\gm = \nu (2-\eta) \; , \qquad \dlt = \frac{5-\eta}{1+\eta} 
\qquad ( d = 3) \; .
\ee
The number of components $N$ corresponds to different physical systems. 
Thus $N=0$ corresponds to dilute polymer solutions, $N = 1$, to the Ising 
universality class, $N=2$, to superfluids and the so-called $XY$ magnetic models, 
$N=3$, to the Heisenberg universality class, and $N=4$, to some models of quantum 
field theory. Formally, it is admissible to study arbitrary $N$. 

In the case of $N=-2$, the critical exponents for any $d$ are known exactly:
$$
\al= \frac{1}{2} \; , \qquad \bt = \frac{1}{4} \; , \qquad \gm = 1 \; ,
$$
\be
\label{293}
\dlt = 5 \; , \qquad \eta = 0 \; , \qquad \nu =\frac{1}{2} 
\qquad ( N = - 2) \;  .
\ee
For the limit $N\ra\infty$, the exact exponents also are available: 
$$
\al = \frac{d-4}{d-2} \; , \qquad \bt = \frac{1}{2} \; , \qquad 
\gm = \frac{2}{d-2} \; , \qquad \dlt = \frac{d+4}{d-2} \; , 
$$
\be
\label{294}
\eta = 0 \; , \qquad \nu =\frac{1}{d-2} \; ,\qquad \om = 4 - d 
\qquad ( N \ra \infty) \;   .
\ee
The latter for $d=3$ reduce to
$$
\al = -1 \; , \qquad \bt = \frac{1}{2} \; , \qquad \gm = 2 \; , 
\qquad \dlt = 5 \; , 
$$
\be
\label{295}
\eta = 0 \; , \qquad \nu = 1 \; , \qquad \om = 1 
\qquad ( d = 3 \;, ~N \ra \infty) \;  .
\ee

The epsilon expansion results in the series
\be
\label{296}
  f_k(\ep) = \sum_{n=0}^k c_n \ep^n \qquad ( \ep \ra 0 ) 
\ee
obtained for $\ep\ra 0$, while at the end we have to set $\ep= 1$. Direct 
substitution of $\varepsilon = 1$ in the series (\ref{296}) leads to the 
values having little to do with real exponents. These series require to define 
their effective sums, which we accomplish by means of the self-similar factor 
approximants 
\be
\label{297}
 f_k^*(\ep) = f_0(\ep) \prod_{i=1}^{N_k} ( 1 + A_i \ep )^{n_i} \;  .
\ee
Then we set $\varepsilon = 1$ and define the final answer as the half sum of 
the last two factor approximants $f_k(1)$ and $f_{k-1}(1)$. 

Let us first illustrate the procedure for the $O(1)$ field theory of the Ising
universality class, where there exist the most accurate numerical calculations 
of the exponents, obtained by Monte Carlo simulations 
\cite{Pelissetto_110,Deng_111,Campostrini_112,Hasenbusch_113,Ferrenberg_114}. 
The epsilon expansions for $\eta$, $\nu^{-1}$, and $\omega$ can be written 
\cite{Kleinert_115} as
$$
\eta \simeq 0.0185185 \ep^2 + 0.01869 \ep^3 - 0.00832877 \ep^4 +
0.0256565 \ep^5 \; ,
$$
$$
\nu^{-1} \simeq 2 - 0.333333\ep - 0.117284 \ep^2 + 0.124527 \ep^3 - 
0.30685 \ep^4 - 0.95124 \ep^5 \; ,
$$
\be
\label{298} 
\om \simeq \ep  - 0.62963 \ep^2 + 1.61822 \ep^3 - 5.23514 \ep^4 + 
20.7498 \ep^5 \;  .
\ee
If we set here $\ep=1$, we get senseless values $\eta=0.0545$, $\nu=2.4049$
and $\om=17.5033$. However by means of the self-similar factor approximants 
we obtain the results shown in Table 3, which are in good agreement with 
Monte Carlo simulations
\cite{Pelissetto_110,Deng_111,Campostrini_112,Hasenbusch_113,Ferrenberg_114}.  
 
The use of the self-similar factor approximants can be extended to the 
calculation of the critical exponents for the arbitrary number of components 
$N$ of the $O(N)$ symmetric $\varphi^4$ field theory in $d=3$. In the general 
case, the epsilon expansions \cite{Kleinert_115} read as
$$
\eta \simeq \frac{(N + 2)\ep^2}{2(N+8)^2}
\left \{  1 + \frac{\ep}{4(N + 8)^2}\;[-N^2 + 56N + 272] -
\right.
$$
$$
-\; \frac{\ep^2}{16(N + 8)^4}\; \left [
5N^4 + 230N^3 - 1124N^2 - 17920N - 46144 + 384\zeta(3)(N + 8)(5N + 22)
\right ] - 
$$
$$
-\; \frac{\ep^3}{64(N + 8)^6}\;\left [ 13N^6 + 946N^5 + 27620N^4 + 121472
N^3 - 262528N^2 - 2912768N - 5655552 - \right.
$$
$$
- 16\zeta(3)(N + 8)\left (N^5 + 10N^4 + 1220N^3 - 1136N^2 - 68672N - 171264
\right ) + 
$$
$$
\left. \left. 
+ 1152\zeta(4)(N + 8)^3 (5N + 22) - 
5120\zeta(5)(N + 8)^2 (2N^2 + 55N + 186) \right ]\right \}  \; ,
$$
\vskip 2mm
$$
\nu^{-1} \simeq 2 + \frac{(N+2)\ep}{N+8} \left\{ -1 -\; 
\frac{\ep}{2(N+8)^2}\;[13N+44] + \right.
$$
$$
+ \frac{\ep^2}{8(N+8)^4}\; \left [ 3N^3-452N^2-2672N - 5312 + 
96\zeta(3)(N+8)(5N+22)\right ] +
$$
$$
+ \frac{\ep^3}{32(N+8)^6}\;\left [ 3N^5+398N^4-12900N^3-81552N^2-
219968N - 357120 + \right.
$$
$$
+ 16\zeta(3)(N+8)\left (3N^4-194N^3+148N^2+9472N+19488\right ) 
+  288 \zeta(4)(N+8)^3(5N+22) - 
$$
$$
\left. - 1280\zeta(5)(N+8)^2\left (2N^2+55N+186\right )\right ] +
$$
$$
+  \frac{\ep^4}{128(N+8)^8}\; \left [3N^7 - 1198N^6 - 27484N^5 - 
1055344N^4 - 5242112N^3 - 5256704N^2 + \right.
$$
$$
+ 6999040N-626688 
- 16\zeta(3)(N+8)\left (13N^6 -310N^5 +19004N^4+ 102400N^3 - 
381536N^2 - \right.
$$
$$
- 2792576N - 4240640) 
- 1024\zeta^2(3)(N+8)^2 \left (2N^4 + 18N^3 + 981N^2 + 6994N + 11688
\right ) +
$$
$$
+ 48\zeta(4)(N+8)^3 \left (3N^4 - 194N^3 + 148N^2 +9472N + 19488
\right ) +
$$
$$
+ 256 \zeta(5)(N+8)^2 \left ( 155N^4 + 3026N^3 + 989N^2 - 66018N - 
130608 \right ) -
$$
$$
\left. \left.
- 6400\zeta(6)(N+8)^4\left (2N^2+55N +186\right ) + 
56448\zeta(7)(N+8)^3\left (14N^2  +189N + 256\right )\right ]
 \right \} \; ,
$$
\vskip 2mm
$$
\om \simeq \ep - \frac{3\ep^2}{(N+8)^2}\; [3N + 14] +
$$
$$
+ \frac{\ep^3}{4(N+8)^4}\; \left [ 33N^3 + 538 N^2 + 4288N + 9568 + 
96\zeta(3)(N+8)(5N + 22)\right ] +
$$
$$
+ \frac{\ep^4}{16(N+8)^6}\; \left[ 5N^5 - 1488N^4 - 46616N^3 - 
419528N^2 - \right.
$$
$$
- 1750080N - 2599552 - 96\zeta(3)(N+8)\left (63N^3 + 548N^2 + 1916N + 
3872\right ) +
$$
$$
\left. + 288\zeta(4)(N+8)^3(5N+22) - 1920\zeta(5)(N+8)^2\left (2N^2 + 
55N + 186\right ) \right ] +
$$
$$
+ \frac{\ep^5}{64(N+8)^8}\;\left [ 13N^7 + 7196N^6 + 240328N^5 + 
3760776N^4 + \right.
$$
$$
+ 38877056N^3 + 223778048N^2 + 660389888N + 752420864 -
$$
$$
-  16\zeta(3)(N+8)\left ( 9N^6 - 1104N^5 - 11648N^4 - 243864N^3 - 
2413248N^2 - 9603328N - 14734080 \right ) -
$$
$$
-  768\zeta^2(3)(N+8)^2\left ( 6N^4 + 107N^3 + 1826N^2 + 9008N + 
8736 \right ) -
$$
$$
-  288 \zeta(4)(N+8)^3\left (63N^3 + 548N^2 + 1916N + 3872
\right ) +
$$
$$
+  256\zeta(5)(N+8)^2\left ( 305N^4 + 7386N^3 + 45654N^2 + 143212N + 
226992\right ) -
$$
$$
- 9600\zeta(6)(N+8)^4\left (2N^5 +55N + 186\right ) +
$$
\be
\label{299}
\left.  +
112896\zeta(7)(N+8)^3\left (14N^2 + 189N + 256\right )\right ] \; .
\ee

Summing these series by means of the self-similar factor approximants 
\cite{Yukalov_116,Yukalov_117}, we obtain the exponents presented in Table 4. 
The found values of the exponents are in good agreement with experimental data 
as well as with the results of numerical methods, such as Pad\'{e}-Borel 
summation and Monte Carlo simulations. It is important to stress that when 
the exact values of the exponents are known (for $N = - 2$ and $N \ra \infty$), 
the self-similar approximants automatically reproduce these exact data.

\section{Conclusion}

In this review, we have presented the basic ideas of the approach allowing 
for obtaining senseful results from divergent asymptotic series typical of 
asymptotic perturbation theory. The pivotal points of the approach can be 
emphasized as follows. 

\vskip 2mm
(i) The implantation of control functions in the calculational procedure, 
treating perturbation theory as optimal control theory. Control functions are 
defined by optimization conditions so that to control the convergence of the 
sequence of optimized approximants. The optimization conditions are derived 
from the Cauchy criterion of sequence convergence. The resulting optimized 
perturbation theory provides good accuracy even for very short series of just 
a few terms and makes it possible to extrapolate the validity of perturbation 
theory to arbitrary values of variables, including the limit to infinity.  

\vskip 2mm
(ii) Reformulation of perturbation theory to the language of dynamical theory, 
handling the motion from one approximation term to another as the motion 
in discrete time played by the approximation order. Then the approximation 
sequence is bijective to the trajectory of the effective dynamical system, 
and the sequence limit is equivalent to the trajectory fixed point. The motion 
near the fixed point enjoys the property of functional self-similarity. The 
approximation dynamical system in discrete time is called cascade. The 
approximation cascade can be embedded into a dynamical system in continuous 
time termed approximation flow. The representation in the language of dynamical 
theory allows us to improve the accuracy of optimized perturbation theory, to 
study the procedure stability, and to select the best initial approximation. 

\vskip 2mm
(iii) Introduction of control functions by means of a fractal transformation 
of asymptotic series, which results in the derivation of several types of 
self-similar approximants. These approximants combine the simplicity of their 
use with good accuracy. They can be employed for the problem of interpolation 
as well as extrapolation.           

\vskip 2mm

The application of the described methods is illustrated by several examples 
demonstrating the efficiency of the approach. 

\vskip 3mm

\begin{table}
\renewcommand{\arraystretch}{1.2}
\centering
\caption{Coefficients $a_n$ of the loop expansion $c_1(x)$ for the number 
of components $N$.}
\vskip 2mm
\label{Table1}    
\begin{tabular}{|c|c|c|c|c|c|} \hline
$N$   &         0    &       1      &    2          &    3    &    4     \\ \hline
$a_1$ &    0.111643  &    0.111643  &    0.111643   &    0.111643   &    0.111643 \\ \hline
$a_2$ & $-$0.0264412 & $-$0.0198309 & $-$0.0165258  & $-$0.0145427  & $-$0.0132206 \\ \hline
$a_3$ &    0.0086215 &    0.00480687&    0.00330574 &    0.00253504 &    0.0020754 \\ \hline
$a_4$ & $-$0.0034786 & $-$0.00143209& $-$0.000807353& $-$0.000536123& $-$0.000392939 \\ \hline
$a_5$ &    0.00164029&    0.00049561&    0.000227835&    0.000130398&    0.0000852025 \\ \hline
\end{tabular}
\end{table}

\begin{table}
\renewcommand{\arraystretch}{1.2}
\centering
\caption{Critical temperature shift obtained using self-similar factor 
approximants, as compared with Monte Carlo simulations.}
\vskip 2mm
\label{Table2}    
\begin{tabular}{|c|c|cc|} \hline
$N$  &         $c_1$      &  $Monte\; Carlo$  &                   \\ \hline
0    &    0.77$\pm$ 0.03  &                   &                    \\ \hline
1    &    1.06$\pm$ 0.05  &  1.09 $\pm$ 0.09  & \cite{Sun_109}    \\ \hline
2    &    1.29 $\pm$ 0.07 &  1.29 $\pm$ 0.05  & \cite{Kashurnikov_106}   \\ 
     &                    &  1.32 $\pm$ 0.02  & \cite{Arnold_107,Arnold_108} \\ \hline
3    &    1.46 $\pm$ 0.08 &                   &                  \\ \hline
4    &    1.60 $\pm$ 0.09 &  1.60 $\pm$ 0.10  & \cite{Sun_109}   \\ \hline
\end{tabular}
\end{table}

\begin{table}
\renewcommand{\arraystretch}{1.2}
\centering
\caption{Critical exponents for the $O(1)$-symmetric $\varphi^4$ field 
theory of the Ising universality class calculated using self-similar factor 
approximants, as compared with Monte Carlo simulations. }
\vskip 2mm
\label{Table 3}    
\begin{tabular}{|c|c|c|} \hline
         & $Factor\; Approximants $ &  $Monte\; Carlo$  \\ \hline
$\al$    &        0.10645           &    0.11026        \\ \hline
$\bt$    &        0.32619           &    0.32630        \\ \hline
$\gm$    &        1.24117           &    1.23708        \\ \hline
$\dlt$   &        4.80502           &    4.79091        \\ \hline
$\eta$   &        0.03359           &    0.03611        \\ \hline
$\nu$    &        0.63118           &    0.62991        \\ \hline
$\om$    &        0.78755           &    0.83000        \\ \hline
\end{tabular}
\end{table}

\begin{table}
\centering
\caption{Critical exponents for the $O(N)$-symmetric $\varphi^4$ field 
theory obtained by the summation of $\varepsilon$ expansions using self-similar 
factor approximants.}
\vskip 2mm
\label{Table4}
\renewcommand{\arraystretch}{1.2}
\begin{tabular}{|c|c|c|c|c|c|c|c|} \hline
$N$&   $\al$     & $\bt$   &   $\gm$ & $\dlt$ & $\eta$  & $\nu$   & $\om$ \\ \hline
-2 &    0.5      &  0.25   &  1      & 5      & 0       & 0.5     & 0.79838 \\
-1 &    0.36612  & 0.27742 & 1.0791  & 4.8897 & 0.01874 & 0.54463 & 0.79380  \\
 0 &    0.23466  & 0.30268 & 1.1600  & 4.8323 & 0.02875 & 0.58845 & 0.79048 \\
 1 &    0.10645  & 0.32619 & 1.2412  & 4.8050 & 0.03359 & 0.63118 & 0.78755 \\
 2 &   -0.01650  & 0.34799 & 1.3205  & 4.7947 & 0.03542 & 0.67217 & 0.78763 \\
 3 &   -0.13202  & 0.36797 & 1.3961  & 4.7940 & 0.03556 & 0.71068 & 0.78904 \\
 4 &   -0.23835  & 0.38603 & 1.4663  & 4.7985 & 0.03476 & 0.74612 & 0.79133 \\
 5 &   -0.33436  & 0.40208 & 1.5302  & 4.8057 & 0.03347 & 0.77812 & 0.79419 \\
 6 &   -0.41963  & 0.41616 & 1.5873  & 4.8142 & 0.03197 & 0.80654 & 0.79747 \\
 7 &   -0.49436  & 0.42836 & 1.6376  & 4.8231 & 0.03038 & 0.83145 & 0.80108 \\
 8 &   -0.55920  & 0.43882 & 1.6816  & 4.8320 & 0.02881 & 0.85307 & 0.80503 \\
 9 &   -0.61506  & 0.44774 & 1.7196  & 4.8406 & 0.02729 & 0.87169 & 0.80935 \\
10 &   -0.66297  & 0.45530 & 1.7524  & 4.8489 & 0.02584 & 0.88766 & 0.81408 \\
50 &   -0.98353  & 0.50113 & 1.9813  & 4.9537 & 0.00779 & 0.99451 & 0.93176 \\
100 &   -0.93643 & 0.49001 & 1.9564  & 4.9926 & 0.00123 & 0.97881 & 0.97201 \\
1000 &  -0.99528 & 0.49933 & 1.9966  & 4.9986 & 0.00023 & 0.99842 & 0.99807  \\
10000 & -0.99952 & 0.49993 & 1.9997  & 4.9999 & 0.00002 & 0.99984 & 0.99979  \\
$\infty$ &  -1   &  0.5    &  2      &  5     &  0      & 1       &  1 \\ \hline
\end{tabular}
\end{table}

\end{document}